\documentclass[11pt, a4paper]{article}
\usepackage{jheparxiv}
\usepackage[utf8]{inputenc}
\usepackage{amsmath}
\usepackage{amsfonts}
\usepackage{amssymb}
\usepackage{latexsym}
\usepackage{mathrsfs}
\usepackage{braket}		
\usepackage{graphicx}
\usepackage{color}
\usepackage{xcolor}
\usepackage{slashed}
\usepackage{twistor}
\usepackage[all]{xy}

\newcommand{\sa}{\mathsf{a}}

\newcommand{\scG}{\mathscr{G}}
\newcommand{\ms}{\mathsf{s}}
\newcommand{\st}{\mathsf{t}}
\newcommand{\su}{\mathsf{u}}

\renewcommand{\d}{\mathrm{d}}
\renewcommand{\sb}{\mathsf{b}}
\renewcommand{\sc}{\mathsf{c}}

\subheader{\hfill \texttt{IMPERIAL-TP-TA-2018-04}}

\title{Plane wave backgrounds and colour-kinematics duality}

\author[a]{Tim Adamo,}
\author[b]{Eduardo Casali,}
\author[c]{Lionel Mason}
\author[c]{\& Stefan Nekovar}

\affiliation[a]{Theoretical Physics Group, Blackett Laboratory \\
        Imperial College London, SW7 2AZ, United Kingdom}

\affiliation[b]{Center for Quantum Mathematics and Physics (QMAP) and\\
Department of Physics, University of California, Davis, CA 95616 USA}
        
\affiliation[c]{The Mathematical Institute \\
        University of Oxford, Woodstock Road, OX2 6GG, United Kingdom}

\emailAdd{t.adamo@imperial.ac.uk}
\emailAdd{ecasali@ucdavis.edu}
\emailAdd{[lmason,nekovar]@maths.ox.ac.uk}

\abstract{We obtain the detailed Feynman rules for perturbative gauge theory on a fixed Yang-Mills plane wave background. Using these rules, the tree-level 4-point gluon amplitude is computed and some 1-loop Feynman diagrams are considered. As an application, we test the extent to which  colour-kinematics duality, the relation between the colour and kinematic constituents of the amplitude, holds on the plane wave background. Although the duality is obstructed, the obstruction has an interesting constrained structure. This plane wave version of colour-kinematics duality reduces on a flat background to the well-known identities underpinning the BCJ relations for colour-ordered partial amplitudes, and constrains representations of tree-level amplitudes beyond 4-points.}

\begin{document}

\maketitle

\section{Introduction}

Recent years have seen the discovery of beautiful structures in perturbative quantum field theory (QFT) that are invisible from the perspective of traditional, Lagrangian based perturbation theory.  A notable  example  is  \emph{colour-kinematics duality}~\cite{Bern:2008qj}, a relationship between the colour structures and kinematic numerators of gauge theory amplitudes  computed perturbatively around flat space. In pure Yang-Mills theory, this duality can be stated as follows: we suppose that the $n$-point, $L$-loop gluon scattering amplitude integrand can be written in the form
\be\label{in1}
A^{(L)}_{n}=\delta^{d}\!\left(\sum_{i=1}^{n}k_{i}\right)\,\int \frac{\d^{Ld}\ell}{(2\,\pi)^{Ld}} \sum_{\Gamma\in\mathrm{cubic}} \frac{c_{\Gamma}\,N_{\Gamma}}{D_\Gamma}\,,
\ee
where $\{k_{i}\}$ are the on-shell external momenta, the sum is over cubic graphs, $\{c_\Gamma\}$ are colour factors (built from the structure constants of the gauge group), $\{D_\Gamma\}$ are scalar propagators (built from the external and loop momenta), and $\{N_\Gamma\}$ are numerators built from the kinematic data. Due to the Jacobi identity, there are linear relations between the colour factors of the form $c_{\alpha}-c_\beta+c_\gamma=0$. Colour-kinematics duality is the statement that there exists a representation of the amplitude \eqref{in1} for which the kinematic numerators obey the \emph{same} linear relations: $N_{\alpha}-N_{\beta}+N_{\gamma}=0$.  Such a representation is said to be a BCJ representation.

It is a theorem that colour-kinematics duality holds at tree-level~\cite{BjerrumBohr:2009rd,Stieberger:2009hq,BjerrumBohr:2010zs,Feng:2010my,Tye:2010dd}.  Furthermore, such BCJ representations have been constructed for supersymmetric theories at 4-points through to 4-loops~\cite{Bern:2012uf,Bern:2012cd,Bern:2013uka,Bern:2014sna}, with generalisations holding at 5-loops~\cite{Bern:2017yxu,Bern:2017ucb}. The duality underpins the Bern-Carrasco-Johansson (BCJ) relations between colour-ordered partial amplitudes~\cite{Bern:2008qj}, which (along with the photon decoupling identity and the Kleiss-Kuijf relations) reduce the number of independent partial amplitudes at tree-level to $(n-3)!$. The power of the duality lies in its relationship with \emph{double copy}~\cite{Bern:2008qj,Bern:2010ue,Bern:2010yg}, which enables the calculation of the gravitational integrands for amplitudes once a BCJ representation of the corresponding gauge theory amplitude has been found. Indeed, for colour-kinematics dual numerators, the gravitational amplitude is given by simply replacing $c_\Gamma\rightarrow N_\Gamma$ in \eqref{in1}. This has enabled computations of supergravity scattering amplitudes to \emph{five} loops~\cite{Bern:2017ucb,Bern:2018jmv}; calculations that would have never been possible with conventional perturbative methods.

Despite its utility, the origins and robustness of colour-kinematics duality and double copy (as well as a host of other novel structures in perturbative QFT) remain mysterious. At tree-level, colour-kinematics duality and the double copy can be understood as the field theory limit of the KLT relations~\cite{Kawai:1985xq} between open and closed string amplitudes, but at loop level their stringy origin is not clear. Optimistically, one hopes that colour-kinematics duality and the double copy are general properties of perturbative gauge theory and gravity; however, if this is true they should hold for perturbation theory on \emph{any} background.

In~\cite{Adamo:2017nia}, we showed that double copy between gauge theory and gravitational 3-point tree amplitudes holds on \emph{plane wave} backgrounds. These are the simplest curved backgrounds which admit a well-posed scattering problem, and are also universal in the sense that any metric~\cite{Penrose:1976} or (free) gauge field~\cite{Ritus:1985,Gueven:2000ru} looks like a plane wave in the neighborhood of a null geodesic. While the fact that double copy holds even at 3-points on the plane wave background is highly non-trivial (the amplitudes encode background-dependent tail effects), a more substantial  test of double copy's robustness will be at 4-points. But in order to perform this test, one must first compute the 4-point, tree-level gluon amplitude on a gauge theory plane wave background.  One can also ask whether some analogue of colour-kinematics holds at that stage.

\medskip

In this paper, we develop perturbative Yang-Mills theory on a plane wave background. The first goal is to compute the 4-point gluon amplitude to provide `theoretical data' for use in future tests of double copy, as well as to test whether there is some remnant of colour-kinematics duality which survives on the plane wave background. After a brief review of the gauge theory plane waves of interest, we determine the Feynman rules of Yang-Mills theory on this perturbative background in section~\ref{PWBG}. The resulting expressions for propagators and vertices (in Feynman-'t Hooft gauge) have not, to our knowledge, appeared before in the literature, although they generalize analogous structures arising in the study of strong field QED on plane wave backgrounds (see~\cite{DiPiazza:2011tq,King:2015tba,Seipt:2017ckc} for recent reviews and references therein).

These Feynman rules are used to calculate the tree-level 4-gluon scattering amplitude on a plane wave background in section~\ref{TreeAmps}, in addition to reproducing the results from~\cite{Adamo:2017nia} for the 3-gluon tree-amplitude. In section~\ref{CKDual}, we show how this 4-point amplitude can be cast into a colour-kinematics form, and demonstrate that although color-kinematics duality is obstructed on a plane wave background, the obstruction is highly structured. Let $\{c_{\ms},c_{\st},c_{\su}\}$ be the colour structures for the $\ms$-, $\st$- and $\su$- channels, respectively. These obey a Jacobi identity $c_{\ms}-c_{\st}+c_{\su}=0$; we give a  plane wave analogue of colour-kinematics duality in the form
\be\label{intropw_ck}
\sigma\left(n_{\ms}-n_{\st}+n_{\su}\right)\simeq 0\,,
\ee
where $\{n_{\ms},n_{\st},n_{\su}\}$ are certain kinematic tree-level integrands associated to each channel, $\sigma$ is a projection map and `$\simeq$' an equivalence relation up to a particular class of obstructions, proportional to the background gauge field. Remarkably, this relation seems to constrain representations of tree-level scattering in the same way as usual colour-kinematics duality, and reduces to the kinematic Jacobi identity $N_{\ms}-N_{\st}+N_{\su}=0$ in the flat background limit.  

Section~\ref{FD} concludes with a discussion of future directions, including the study of double copy, higher numbers of external gluons, loops and the prospects for computing these amplitudes with \emph{ambitwistor strings}, a class of  worldsheet models.


\section{Perturbative Gauge Theory on a Plane Wave}
\label{PWBG}

We construct perturbative Yang-Mills theory on a non-trivial plane wave background gauge field. In this section, we review features of plane wave backgrounds and go on to derive the Feynman rules (free fields, vertices and propagator) for gauge theory on this background. 

\subsection{The plane wave background}

We take a vacuum plane wave gauge potential on Minkowski space-time to be a solution to the (source-free) Yang-Mills equations that has $(2d-3)$ symmetries, which form a Heisenberg algebra with centre given by a null translation~\cite{Wolkow:1935zz,Schwinger:1951nm,Coleman:1977ps,Heinzl:2017blq,Adamo:2017nia}. We restrict to gauge fields that are valued in the Cartan of the gauge group. In light-cone (null) coordinates for which the Minkowski metric reads
\be\label{Mink}
\d s^2= \d X_{\mu}\,\d X^{\mu}= 2\, \d x^{+}\,\d x^{-} - \d x_{a}\,\d x^{a}\,,
\ee
for $X^{\mu}=(x^{+},x^a,x^-)$, $a=1,\ldots,d-2$, a plane wave gauge field is
\be\label{gpw1}
A=-A_{a}(x^-)\,\d x^{a}\,,
\ee
where the coefficient functions $A_{a}$ are valued in the Cartan of the gauge group (and the overall negative sign is for later convenience).

This is easily seen to be a solution to the (Cartan-valued) Yang-Mills equations, and is clearly symmetric under transformations generated by $\partial_{+}$, $\partial_{a}$. There are $(d-2)$ further symmetries of \eqref{gpw1} which are less obvious, generated by
\be\label{gpw2}
\cX^{a}=x^{a}\,\partial_{+}+x^{-}\,\partial^{a} +\int^{x^-}\!\!\d s\,A^{a}(s)\,,
\ee
which correspond to null rotations (cf., \cite{Heinzl:2017zsr,Heinzl:2017blq,Ilderton:2018lsf}). The set of $(2d-3)$ symmetry generators $\{\partial_{+},\partial_{a},\cX^{b}\}$ form an algebra whose only non-trivial commutators are
\be\label{Heis}
\left[\partial_{a},\,\cX^{b}\right]=\delta^{a}_{b}\,\partial_{+}\,,
\ee
which is the required Heisenberg algebra. It is straightforward to check that the generator $\partial_{+}$ is in fact covariantly constant.

The plane wave solution can be understood as a coherent superposition of gluons. A linearised gluon can be represented using the familiar momentum eigenstate $\mathsf{T}^{\sa}\,\epsilon_{\mu}\,\e^{\im k\cdot X}$, where $k^2=0$, $\mathsf{T}^{\sa}$ is a generator of the gauge group and $\epsilon_{\mu}$ is an on-shell polarization. Since the momentum $k_{\mu}$ is null, it defines the light cone direction $x^{-}=k\cdot X$. The on-shell polarization of a gluon in $d$-dimensions has $d-2$ degrees of freedom, labelled with index $a=1,\ldots, d-2$. Aligning the colour vector $\mathsf{T}^{\sa}$ with the Cartan of the gauge group, one obtains \eqref{gpw1} from such  a superposition of such free gluons.

\medskip

\begin{figure}[t]
\centering
\includegraphics[scale=.5]{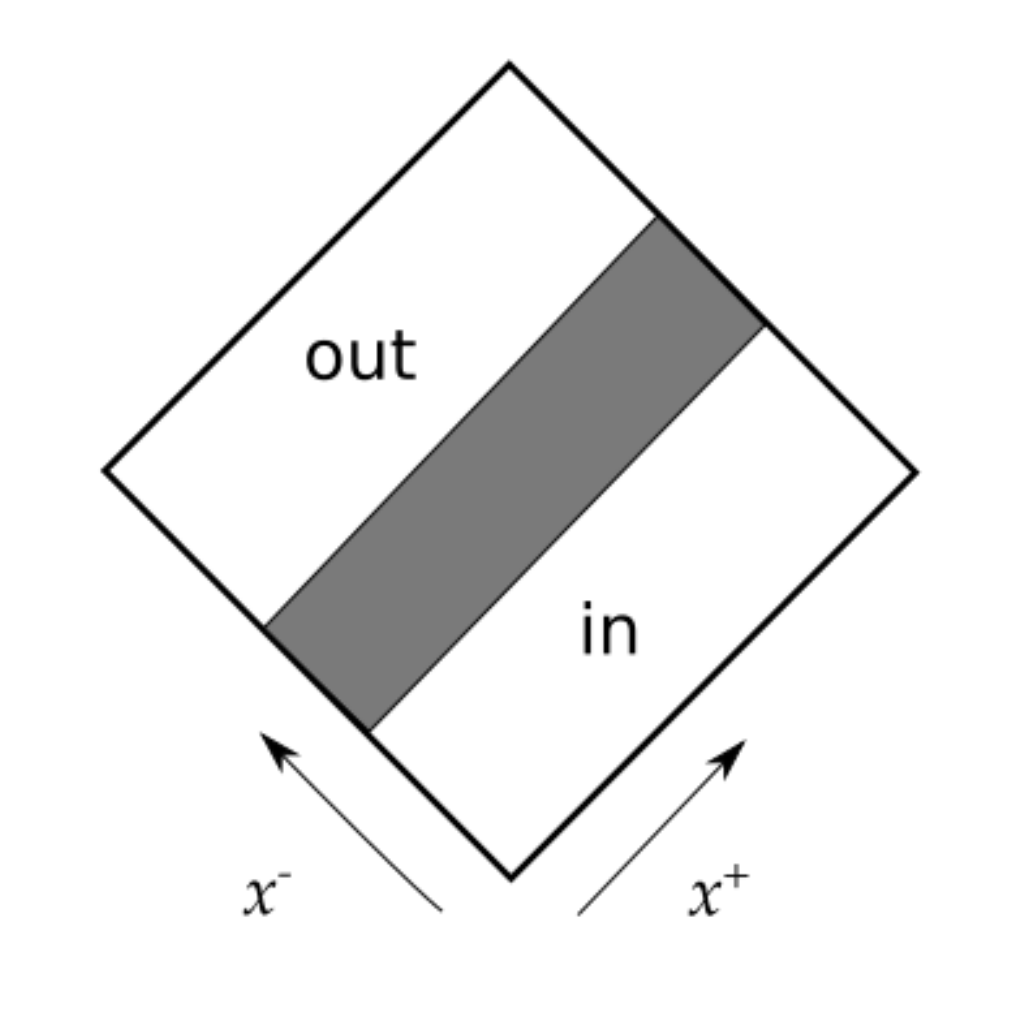}
\caption{The sandwich plane wave with $x^a$-directions suppressed. The field strength $F_{a}=\dot{A}_{a}(x^-)$ is non-zero only in the shaded region; the gauge field is flat on the in- and out-regions.}
\label{SandwichPW}
\end{figure}

The field strength of the plane wave is
\be\label{gpw3}
F=\dot{A}_{a}(x^-)\,\d x^{a}\wedge\d x^{-}\,,
\ee
where $\dot{A}_{a}=\partial_{-}A_{a}$.
By taking $A\rightarrow A +\d (x^{a} A_{a})$, we obtain the alternative representation
\be\label{gpwBrink}
A=x^{a}\,\dot{A}_{a}(x^-)\,\d x^-\,.
\ee
This gauge has the advantage that there is only one non-trivial component of the gauge potential in the $x^-$-direction and it is only non-zero when the curvature is non-zero. For these reasons, we always work in this gauge \eqref{gpwBrink}, which can be thought of as a `Brinkmann' gauge, analogous to the Brinkmann coordinates of a gravitational plane wave (which are global, encode the curvature algebraically and have only one non-trivial metric component~\cite{Brinkmann:1925fr}).

To obtain a well-defined scattering problem, we restrict attention to \emph{sandwich} plane waves~\cite{Bondi:1958aj}, for which the field strength $F_{a}=\dot{A}_{a}$ is compactly supported in some range $x^{-}_{1}\leq x^{-}\leq x^{-}_{2}$ (see Figure \ref{SandwichPW}). 
It can be shown that the S-matrix for gluons on a sandwich plane wave background is well-defined, in the sense that evolution from the in-region to the out-region is unitary and there is no particle creation (in the quadratic theory)~\cite{Schwinger:1951nm,Adamo:2017nia}.


\subsection{Feynman rules}

Perturbative Yang-Mills theory on a background is described by evaluating the interaction terms in the classical Yang-Mills action on a gauge field $\cA=A+a$, where $A$ is the background gauge field and $a$ is the fluctuating field appearing in the path integral measure (cf., \cite{DeWitt:1967ub,tHooft:1975uxh,Boulware:1980av,Abbott:1981ke} or chapter 16.6 of \cite{Peskin:1995ev}). The resulting action has the form
\be\label{YMa1}
S[a;A]=S_{\mathrm{kin}}[a;A] + S_{\mathrm{int}}[a;A]\,,
\ee
where $S_{\mathrm{kin}}$ is quadratic in the fluctuation $a$ while $S_{\mathrm{int}}$ is $O(a^3)$. After gauge fixing, position-space Feynman rules are obtained from this action in the usual way: free fields and propagators are determined by $S_{\mathrm{kin}}$ while the interaction vertices are read off from $S_{\mathrm{int}}$. It will be convenient to work in Feynman-'t Hooft gauge, for which the gauge-fixing term added to the space-time action is 
\be\label{Feyngauge}
S_{\mathrm{GF}}[a;A]=-\frac{1}{2\,g^2}\int \d^{d}X\,\tr \left(D^{\mu}a_{\mu}\right)^2\,,
\ee
for $g$ the Yang-Mills coupling and $D_{\mu}=\partial_{\mu}-\im [A_{\mu},\cdot]$ the covariant derivative with respect to the background field.

\subsubsection{Free fields}

Solutions to the free equations of motion in the background serve as on-shell external states for insertion on external lines in Feynman diagrams via the LSZ-reduction formula. In Feynman-'t Hooft gauge, the kinetic part of the action \eqref{YMa1} reads
\be\label{YMkin1}
S_{\mathrm{kin}}[a;A]= -\frac{1}{g^2}\int \d^{d}X\,\tr\left(D_{[\mu}a_{\nu]}\,D^{[\mu}a^{\nu]}+\frac{1}{2}F^{\mu\nu}\,[a_{\mu},a_{\nu}]+\frac{1}{2} (D^{\mu}a_{\mu})^2\right)\,,
\ee
where $F_{\mu\nu}$ is the field strength of the background field and $D_{[\mu}a_{\nu]}=\frac{1}{2}(D_{\mu}a_{\nu}-D_{\nu}a_{\mu})$. This gives the linearised equation of motion for the gauge field fluctuation $a_{\mu}$ as
\be\label{YMkin2}
D_{\nu}D^{\nu}\,a_{\mu}+2\im\,[F_{\mu\nu},\,a^{\nu}]=0\,.
\ee
This simplifies further when the background $A$ is a plane wave valued in the Cartan of the gauge group. The  commutators involving $A$ can then be expressed in terms of the charge $e$ of the fluctuation with respect to the background. In general, this charge is the root $\lambda$ when the fluctuation $a$ takes values in the corresponding  root space:
Let $\mathsf{T}^{\sa}$ be an element of the root space for $\lambda$ in  the Lie algebra of the gauge group, and $\mathsf{t}^{\mathsf{i}}$ generators of the Cartan subalgebra; then the charge is defined by $(e^{\sa})^{\mathsf{i}}:=[\mathsf{t}^{\mathsf{i}},\mathsf{T}^{\sa}]=\lambda(\mathsf{t}^{\mathsf{i}})$. Since $A=\mathsf{t}^{\mathsf{i}} A^{\mathsf{i}}$, it follows that $[A_{\mu},a_{\nu}]=e A_{\mu} a_{\nu}$, where contractions in the Cartan subalgebra are implicit, i.e., we define $eA:=e^{\mathsf{i}}A^{\mathsf{i}}$. 

Inserting the explicit form of the plane wave background into \eqref{YMkin2}, the free equation of motion becomes:
\be\label{pwkin1}
(2\,\partial_{+}\partial_{-}-\partial_{a}\partial^{a}-2\im\,e\,x^{a}\dot{A}_{a} \partial_{+}) a_{\mu} +2\im\,e\,(\delta_{\mu}^{a}\dot{A}_{a}\,a_{+}+\delta_{\mu}^{-}\dot{A}^{a}\,a_{a})=0\,.
\ee
The first set of bracketed terms in this equation is the charged scalar wave operator
\be
\Box_{A}:=D_\mu D^\mu=(2\,\partial_{+}\partial_{-}-\partial_{a}\partial^{a}-2\im\,e\,x^{a}\dot{A}_{a} \partial_{+})\,.
\ee
  It will be convenient for constructing propagators to have solutions to the massive charged scalar wave equation in a plane wave background  
  \be
  (\Box_{A}+m^2)\Phi=0
  \ee which are given in an extension of ~\cite{Adamo:2017nia} to non-zero mass by
\be\label{cswe}
\Phi=\mathsf{T}^{\sa}\,\e^{\im\,\phi}\,,
\ee
where $\phi$ is a solution to the (gauge-covariant) massive Hamilton-Jacobi equations\footnote{Tensors in the $(d-2)$-dimensional transverse directions are denoted using boldface for index-free expressions (e.g., $k_{a}\leftrightarrow\bk$).}:
\be\label{phi}
\phi:=k_{+}\,x^{+}+(k_{a}+eA_{a})\,x^{a}+\frac{f(x^-)}{2\,k_{+}}\,, \qquad f(x^{-}):=\int^{x^-}\!\!m^2+(\bk +e \mathbf{A}(s))^2\, \d s\,.
\ee
The constant momentum components $(k_{+},k_a,k_-)$ are the degrees of freedom of a  $d$-vector with norm $m^2$ 
\be
k^2:=2k_+k_--\bk_a\bk^a=m^2\, .
\ee
and in the flat background limit ($A_{a}\rightarrow \mathrm{const.}$)  $\phi\rightarrow k\cdot X$, where $k_{\mu}=(k_{+},k_{a}, k_-)$ is an on-shell momentum of mass $m$ in light cone coordinates.

Solutions to the charged scalar wave equation in an abelian plane wave background have been intensively studied in the context of strong field QED for decades, and are often known in the laser physics literature as `Volkov solutions' (e.g., \cite{Wolkow:1935zz} and \cite{Seipt:2017ckc} for a recent review). The only difference between \eqref{cswe} and these Volkov solutions is a gauge transformation and the proviso that $e$ is root-valued.

\medskip

Armed with the charged scalar solution \eqref{cswe}, it is straightforward to solve the gluon equation of motion \eqref{pwkin1}.  We extend non-zero mass (again for the construction of propagators) replacing $\Box_A$ by $\Box_A+m^2$ in \eqref{pwkin1} using the ansatz $a_{\mu}(X)=P_{\mu}(x^{-})\,\Phi(X)$. This gives a system of coupled ODEs for the components of the unspecified vector $P_{\mu}(x^{-})$:
\be\label{pweom1}
\dot{P}_{+}=0\,, \qquad \dot{P}_{a}=\frac{e}{k_+}\,\dot{A}_{a}\,P_{+}\,, \qquad \dot{P}_{-}=\frac{e}{k_+}\,\dot{A}^{a}\,P_{a}\,.
\ee
These are solved by taking
\begin{equation*}
P_{+}(x^-)=c_{+}\,, \qquad P_{a}(x^-)=c_{a}+\frac{c_{+} }{k_+}\, eA_{a}(x^-)\,,
\end{equation*}
\be\label{pweom2}
P_{-}(x^-)=c_- + \frac{c_{a} }{k_+}\, eA^{a}(x^-) + \frac{c_{+} }{2 k_{+}^2} (e\mathbf{A})^2(x^-)\,,
\ee
for $(c_{+},c_{a},c_{-})$ constants of integration. This allows us to write
\be\label{pweom3}
P_{\mu}(x^{-})=P_{\mu}{}^{\nu}(x^{-})\,c_{\nu}=\left(\begin{array}{ccc}
                                                      1 & 0 & 0 \\
                                                      \frac{e A_{a}}{k_+} & \delta_{a}^{b} & 0 \\
                              \frac{ (e\mathbf{A})^2}{2k_+^2} & \frac{e A^{b}}{k_+} & 1
                                                     \end{array}\right) \, \left(\begin{array}{c}
                                                                                  c_{+} \\
                                                                                  c_{b} \\
                                                                                  c_{-}
                                                                                 \end{array}\right)\,,
\ee
with the matrix $P_{\mu}{}^{\nu}$ encoding all of the dependence on the background field.

On shell external particles are now obtained by taking $m=0$.  
However, this appears to lead to solutions which depend on $d$ free parameters, which is two greater than the number of on-shell polarizations for a gluon in $d$-dimensions. This over-counting is remedied by ensuring that the gauge condition $D^{\mu} a_{\mu}=0$ is satisfied by \eqref{pweom3}. For the plane wave background, this condition can be expressed as $K^{\mu} P_{\mu}=0$, where $K_{\mu}$ is the natural on-shell momentum associated with the solution:
\be\label{momentum}
K_{\mu}:=-\im\,\e^{-\im\phi}D_{\mu}\,\e^{\im\phi}=\left(k_{+},\;k_{a}+e\,A_{a}, \; \frac{(\bk+e \mathbf{A})^2}{2\,k_+}\right)\,,
\ee
which now obeys $K^2=0$. This gauge constraint forces
\be\label{pwos}
c_+=0\,, \qquad c_a=\epsilon_a\,, \qquad c_{-}=\frac{\mathbf{\epsilon}\cdot\bk}{k_+}\,,
\ee
so Feynman gauge imposes Lorenz and lightcone gauge on the free fields simultaneously as a result of the highly symmetric nature of the background~\cite{Adamo:2017nia}. 

Thus, free gluon solutions in Feynman gauge on the plane wave background are given by:
\be\label{gluon}
a_{\mu}=\mathsf{T}^{\sa}\,\epsilon_{\mu}\, \e^{\im\,\phi}\,, \qquad \epsilon_{\mu}=P_{\mu}{}^{\nu}\,c_{\nu}\,,
\ee
with the values of $c_{\mu}$ fixed by \eqref{pwos}. Although both the polarization and the momentum are non-trivial functions of the light cone coordinate $x^-$ through their dependence on the background field, they remain on-shell in the usual sense: $K^2=0=K^{\mu}\epsilon_{\mu}$. Furthermore, there are $(d-2)$ degrees of freedom in the polarization, given by the constants $\epsilon_{a}$; these match the $(d-2)$ on-shell polarizations of a gluon in $d$-dimensions.

\medskip

These solutions exhibit  a \emph{memory effect}~\cite{Bieri:2013hqa,Pate:2017vwa}, whereby if $\phi\rightarrow k\cdot X$ in the past in-region ($x^-<x^{-}_{1}$) of a sandwich wave, then in the future the out-region ($x^{-}>x^{-}_{2}$) we will have 
\be\label{mem}
\phi\rightarrow k_{+}\,x^{+}+(\bk_{a}+eA_{a})\,x^{a}+\left(k_-+\frac{2\bk\cdot e\mathbf{A}+(e\mathbf{A})^2}{2k_+}\right) x_-
\ee
with $eA_a =\int_{-\infty}^\infty \dot A_a(s)ds$ is non vanishing in general although  the mass is unchanged.  Thus the momenta receives a kick from the total integral of the  field strength from one  asymptotic  region to the other.

This memory effect \eqref{mem} introduces a subtlety associated with these free fields on the sandwich plane wave background which does not exist in a flat background. This is a functional distinction between `in-states' and `out-states' (cf., \cite{Kibble:1965zza,Gibbons:1975jb,Garriga:1990dp,Dinu:2012tj,Adamo:2017nia}). In particular, a free gluon $a_{\mu}$ can be chosen to have boundary conditions $a_{\mu}=\epsilon_{\mu}\e^{\im\,k\cdot X}$ on the in-region ($x^{-}<x^{-}_{1}$) or the out-region ($x^{-}>x^{-}_{2}$), but \emph{not both}. For simplicity, we insert all in-state gluons on external legs of Feynman diagrams, although the formalism makes it clear how to insert any combination of external in- or out-states.  

\subsubsection{Propagator}

The kinetic portion of the action also defines the position space propagator; on the plane wave background this is a Green's function
\be\label{prop1}
\left(D_{\lambda}D^{\lambda}\delta_{\sigma}^{\mu}+2\im\,e F_{\sigma}{}^{\mu}\right) \scG_{\mu\nu}(X,Y)= \delta^{\sa\sb}\,\eta_{\sigma\nu}\, \delta^{d}(X-Y)\,,
\ee
where $X,Y$ are any two points in Minkowski space (and the trivial colour structure of the propagator is suppressed). Our aim will be to construct the Feynman propagator solution to this equation in Feynman-'t Hooft gauge; this can be done by taking a sum
\be\label{prop2}
\scG^{F}_{\mu\nu}(X,Y)=\im\,\Theta(x^{-}-y^{-})\,\sum_{i} p^{i}_{\mu}(X)\,\bar{p}^{i}_{\nu}(Y) -\im\,\Theta(y^{-}-x^{-})\,\sum_{i} n^{i}_{\mu}(X)\,\bar{n}^{i}_{\nu}(Y)\,,
\ee
where $\{p^{i}_{\mu}\}$ and $\{n^i_{\mu}\}$ are a basis of positive and negative frequency states, respectively. The free fields \eqref{gluon} provide such a basis, and since the Feynman propagator is invariant under any change of basis which is a non-mixing Bogoliubov transformation~\cite{Gibbons:1975jb}, we are free to use either in-states or out-states to construct the propagator.  However, it is easier to use the off-shell states introduced in the previous subsection.

To realize this explicitly, we use the  off-shell version of the function $\phi(X)$, \eqref{phi} 
so that the scalar function $\e^{\im\,\phi_k}$ solves an off-shell charged wave equation
\be\label{os2}
D_{\mu} D^{\mu}\e^{\im\,\phi_k}=(\bk^2-2k_{+}k_{-})\e^{\im \, \phi_k}\,,
\ee
which becomes on-shell only when $k_{-}=\frac{\bk^2}{2k_+}$. The Feynman propagator is then built from a superposition of these off-shell modes:
\be\label{fprop1}
\scG^{F}_{\mu\nu}(X,Y)=\frac{\cN_{(d)}\,\delta^{\sa\sb}}{2\,\pi\,\im}\,\int \frac{\d^{d}k}{k^{2}+\im\,\varepsilon}\,\D_{\mu\nu}(x^{-},y^{-})\,\exp\left[\im\phi_k(X)-\im\phi_{k}(Y)\right]\,,
\ee
where integration is over the $d$ degrees of freedom $\{k_{+},k_{a},k_{-}\}$ of an off-shell momentum, evaluated on the usual Feynman contour in the complex $(k_{-})$ -- plane. The matrix $\D_{\mu\nu}(x^-,y^-)$ is defined by 
\be\label{fprop}
\D_{\mu\nu}(x^{-},y^{-}):= P_{\mu}{}^{\sigma}(x^{-})\,P_{\nu\sigma}(y^{-})=\left(\begin{array}{ccc}
                                                                                0 & 0 & 1 \\
                                                                                0 & -\delta_{ab} & \frac{e\Delta A_{a}}{k_+} \\
                                                                                1 & -\frac{e \Delta A_{b}}{k_+} & \frac{e^2 \Delta \mathbf{A}^{2}}{2k^2_+}
                                                                                                                                             \end{array}\right)\,,
\ee
for $P_{\mu\nu}$ given by \eqref{pweom3}, $\Delta A_{a}=A_{a}(x^-)-A_{a}(y^-)$, and $\cN_{(d)}$ is a dimension-dependent normalisation constant
\be\label{norm}
\cN_{(d)}:=-\frac{\pi\,\im}{(2\,\pi)^d}\,.
\ee
From now on, the trivial colour structure $\delta^{\sa\sb}$ will be left implicit.

We can explicitly perform all but one of the remaining $k$ integrals.  Performing the contour integration in $\d k_{-}$ leaves:
\begin{multline}\label{fprop2}
\scG^{F}_{\mu\nu}(X,Y)= \cN_{(d)}\,\Theta(x^--y^-)\,\int \frac{\d k_{+}\,\d^{d-2}\bk}{k_+}\,\D_{\mu\nu}(x^{-},y^{-})\, \exp\left[\im\phi(X)-\im\phi(Y)\right] \\
-\cN_{(d)}\,\Theta(y^{-}-x^{-})\,\int \frac{\d k_{+}\,\d^{d-2}\bk}{k_+}\,\D_{\mu\nu}(x^{-},y^{-})\, \exp\left[\im\phi(Y)-\im\phi(X)\right]\,,
\end{multline}
where $\phi(X)$ is now the (on-shell) function \eqref{phi}. This now gives the form \eqref{prop2}.

The integrals over $\d^{d-2}\bk$ are Gaussian, and can be done explicitly with the result
\begin{multline}\label{fprop3}
 \scG^{F}_{\mu\nu}(X,Y)=\cN_{(d)}\,(2\pi\im)^{\frac{d-2}{2}}\,\Theta(\Delta x^{-})\int\limits_{0}^{\infty} \frac{\d k_+}{k_+} \left(\frac{k_+}{\Delta x^-}\right)^{\frac{d-2}{2}}\,\D_{\mu\nu}(x^-,y^-)\\
 \times\exp\left[\im\left(k_+\,\Delta x^{+}-\frac{k_+}{2\Delta x^{-}}\,(\Delta\tilde{\mathbf{x}})^2+e\,\Delta(\mathbf{A}\cdot\mathbf{x}) +\frac{e^2}{2\,k_+} \int_{y^-}^{x^{-}}\!\! \d s\,\mathbf{A}^{2}(s)\right)\right] \\
 -\cN_{(d)}\,(2\pi\im)^{\frac{d-2}{2}}\,\Theta(-\Delta x^{-})\int\limits_{0}^{\infty} \frac{\d k_+}{k_+} \left(\frac{k_+}{\Delta x^-}\right)^{\frac{d-2}{2}}\,\D_{\mu\nu}(x^-,y^-)\\
 \times\exp\left[-\im\left(k_+\,\Delta x^{+}-\frac{k_+}{2\Delta x^{-}}\,(\Delta\tilde{\mathbf{x}})^2+e\,\Delta(\mathbf{A}\cdot\mathbf{x}) +\frac{e^2}{2\,k_+} \int_{y^-}^{x^{-}}\!\! \d s\,\mathbf{A}^{2}(s)\right)\right]\,,
\end{multline}
adopting the shorthand $\Delta X^{\mu}=X^{\mu}-Y^{\mu}$ for differences in position and denoting
\be\label{cnote}
\Delta\tilde{x}^{a}:=\Delta x^{a}+\frac{e}{k_+} \int_{y^-}^{x^{-}}\!\!\d s\,A^{a}(s)\,, \qquad \Delta(\mathbf{A}\cdot\mathbf{x}):=A_{a}(x^-) x^{a}-A_{a}(y^-) y^{a}\,.
\ee
For a general plane wave background, the remaining $\d k_+$ integration cannot be performed analytically due to the wave profile; this is a common feature in perturbative QFT on plane wave backgrounds. However, in the flat background limit, it is easy to see that 
\be\label{flatprop}
\lim_{\mathbf{A}\rightarrow\mathrm{const.}} \scG^{F}_{\mu\nu}(X,Y)= -\frac{\Gamma\left(\frac{d-2}{2}\right)\,\im^{d-1}}{4\,\pi^{\frac{d}{2}}}\,\frac{\eta_{\mu\nu}}{(\Delta X^2)^{\frac{d-2}{2}}}\,,
\ee
which is the familiar expression for the position space gluon propagator on a flat background.

\medskip

$\scG^{F}_{\mu\nu}$ is in Feynman-'t Hooft gauge by construction, so it remains to show that it obeys the equation \eqref{prop1} for a gluon propagator. 
Applying the wave operator to \eqref{fprop1} and using the fact that each numerator factor satisfies the off-shell equation, we obtain
\be
\left(D_{\lambda}D^{\lambda}\delta_{\sigma}^{\mu}+2\im\,e F_{\sigma}{}^{\mu}\right) \scG_{\mu\nu}(X,Y)=
\frac{\cN_{(d)}}{2\,\pi\,\im}\,\int \d^{d}k\,\D_{\mu\nu}(x^{-},y^{-})\,\exp\left[\im\phi_k(X)-\im\phi_{k}(Y)\right]\,.
\ee
The argument of the exponential is
$$
\Delta\phi_{k}=k_{+}\,\Delta x^{+}+\Delta((k_{a}+eA_{a})\,x^{a})+k_{-}\Delta\,x^{-}+\frac{1}{2\,k_+}\int^{x^-}_{y^-}\!\!\left(2e\,\bk\cdot\mathbf{A}(s)+e^2\,\mathbf{A}^2(s)\right) \d s
$$
and so we see that the 
 $k_-$ integral can be performed directly to give a factor of $\delta(x^--y^-)$. On the support of this delta function, $D_{\mu\nu}$ reduces to  $\eta_{\mu\nu}$ and the argument of the exponential reduces to
$$
 \Delta\phi_{k}=k_{+}\,( x^{+}-y^+)+k_{a}(x^{a}-y^a)\,,
$$
so 
 the remaining $k$ integrals can be performed to yield the $d-1$ further required delta functions:
\be\label{pp4}
\left(D_{\lambda}D^{\lambda}\delta_{\sigma}^{\mu}+2\im\,e F_{\sigma}{}^{\mu}\right) \scG^{F}_{\mu\nu}(X,Y)=\eta_{\sigma\nu}\,\delta^{d}(\Delta X)\,.
\ee
Thus our expression for $\scG^{F}_{\mu\nu}$ is indeed the gluon propagator in Feynman-'t Hooft gauge.

\medskip

The object $\D_{\mu\nu}(x^-,y^-)$  captures the tensor structure associated with gluon propagation on the plane wave background and reduces to $\eta_{\mu\nu}$ in the flat background limit. If $K_{\mu}$ is the momentum exchanged by the propagator, then
\be\label{pident}
D_{\mu\nu}(x^-,y^-)\,K^{\nu}(y^-)=K_{\mu}(x^-)\,,
\ee
so that  $\D_{\mu\nu}$  propagates the position dependence of the exchanged momentum from one end of the propagator to the other, as is necessary for pure gauge. This identity will be useful later.

\subsubsection{Vertices}

\begin{figure}[t]
\centering
\includegraphics[scale=.6]{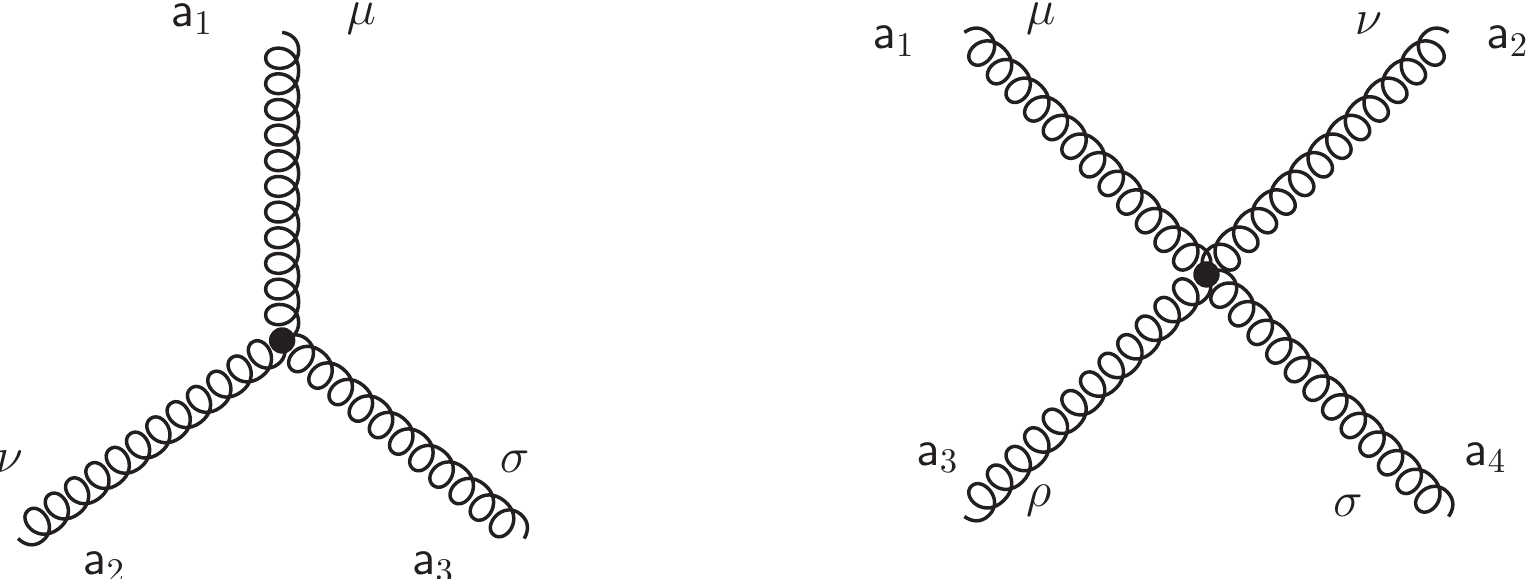}
\caption{The three- and four-point gluon vertices}
\label{Feynrules1}
\end{figure}

The interacting portion of the Yang-Mills action with a background field is
\be\label{Sint}
S_{\mathrm{int}}[a;A]=-\frac{1}{4\,g^2}\int \d^{d}X\,\tr\left(4\,[a_{\mu},\,a_{\nu}]\,D^{[\mu}a^{\nu]}+[a_{\mu},\,a_{\nu}]\,[a^{\mu},\,a^{\nu}]\right)\,.
\ee
As usual, this defines the 3- and 4-point interaction vertices of the theory, which are given in position space by (see Figure \ref{Feynrules1}):
\be\label{3pv}
V^{\mu\nu\sigma}=g\,f^{\sa_{1}\sa_{2}\sa_{3}}\,\int \d^{d}X\,\left(\eta^{\mu\nu}\,(D_{1}-D_2)^{\sigma}+\eta^{\nu\sigma}\,(D_{2}-D_{3})^{\mu}+\eta^{\sigma\mu}\,(D_{3}-D_{1})^{\nu}\right)\,,
\ee
\begin{multline}\label{4pv}
V^{\mu\nu\rho\sigma}=g^{2}\,\int \d^{d}X\,\left[f^{\sa_{1}\sa_{2}\sb}f^{\sa_{3}\sa_{4}\sb}\,(\eta^{\mu\rho}\eta^{\nu\sigma}-\eta^{\mu\sigma}\eta^{\nu\rho}) \right. \\
\left. + f^{\sa_{1}\sa_{3}\sb}f^{\sa_{2}\sa_{4}\sb}\,(\eta^{\mu\nu}\eta^{\rho\sigma}-\eta^{\mu\sigma}\eta^{\nu\rho})+f^{\sa_{1}\sa_{4}\sb}f^{\sa_{2}\sa_{3}\sb}\,(\eta^{\mu\nu}\eta^{\rho\sigma}-\eta^{\mu\rho}\eta^{\nu\sigma})\right]\,,
\end{multline}
where $f^{\sa\sb\sc}$ are the structure constants of the gauge group. Implicit at each vertex is a Kronecker delta conserving the total charge with respect to the background gauge field. In the three-point vertex, the $D_1^\mu$ is understood to act on the first leg coming into the vertex and so on. On an external field the derivative $D_1^\mu$ produces a term with a factor of $K_1^\mu$ but there is an additional term in the $x_-$ direction arising from the derivative of $P_{\mu\nu}$. It is easy to see that the flat limit reproduces the usual momentum space vertices of Yang-Mills theory.

\subsubsection{Ghosts}
 Although our primary concern in this paper will be tree-level, at loops we will need to
 include the Feynman rules for the Fadeev-Popov ghosts that arise from gauge-fixing in non-abelian gauge theories. We include these on the plane wave background to give a complete description of the perturbative theory. The Fadeev-Popov procedure leads to a ghost action on the background
\be\label{ghost1}
S_{\mathrm{ghost}}[c,\bar{c},a;A]=\frac{1}{4\,g^2}\int \d^{d}X\,\tr\left(\bar{c}\,D_{\mu}D^{\mu}\,c+\bar{c}\,D^{\mu}[a_{\mu},c]\right)\,,
\ee
where $c^{\sa}$, $\bar{c}^{\sa}$ are the anti-commuting scalar ghosts, valued in the adjoint of the gauge group. 

\begin{figure}[t]
\centering
\includegraphics[scale=.6]{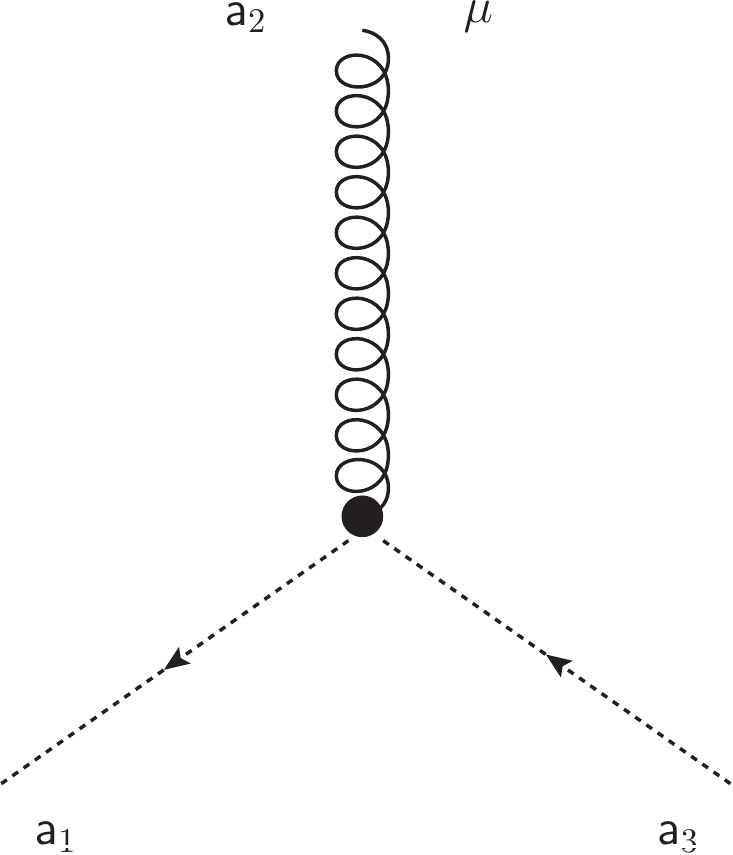}
\caption{The ghost vertex}
\label{Feynrules2}
\end{figure}

The ghosts do not appear on external legs of Feynman diagrams, and the $c-\bar{c}$ propagator and ghost-gluon-ghost vertex are easily expressed using the ingredients already developed. In particular, the ghost propagator is given by
\be\label{ghostp}
\left\la c^{\sa}(X)\,\bar{c}^{\sb}(Y)\right\ra=\frac{\cN_{(d)}\,\delta^{\sa\sb}}{2\,\pi\,\im}\,\oint \frac{\d^{d}k}{k^{2}+\im\,\varepsilon}\,\exp\left[\im\phi_k(X)-\im\phi_{k}(Y)\right]\,,
\ee
while the cubic vertex involving two ghosts and one gluon is (see Figure \ref{Feynrules2})
\be\label{ghostv}
V^{\mu}=g\,f^{\sa_{1}\sa_{2}\sa_3}\,\int \d^{d} X\,D^{\mu}_{1}\,.
\ee
These complete the full set of Feynman rules for gauge theory on the plane wave background.


\section{Tree Amplitudes}
\label{TreeAmps}

Using the Feynman rules, any perturbative calculation on the plane wave background can now be done (at least in principle). To demonstrate this, we consider the examples of the tree-level 3-point and 4-point gluon amplitudes.


\subsection{3-points}

The tree-level 3-point amplitude is given simply by evaluating the cubic vertex \eqref{3pv} on three on-shell states, whose charges obey $e_1+e_2+e_3=0$ (the amplitude vanishes otherwise)~\cite{Adamo:2017nia}. This results in:
\begin{multline}\label{3p1}
\im g\,f^{\sa_{1}\sa_{2}\sa_{3}}\,\int \d^{d}X\,\left(\epsilon_{1}\cdot\epsilon_{2}\,\epsilon_{3}\cdot(K_{1}-K_2)+\epsilon_{2}\cdot\epsilon_{3}\,\epsilon_{1}\cdot(K_{2}-K_{3})+\epsilon_{1}\cdot\epsilon_{3}\,\epsilon_{2}\cdot(K_{3}-K_{1})\right) \\
\times\,\exp\left[\im \sum_{r=1}^{3} \phi_{r}\right]\,.
\end{multline}
The position space integrals over $\d x^{+}$ and $\d^{d-2}\mathbf{x}$ can be performed explicitly, resulting in momentum conserving delta functions. The final integration, over $\d x^{-}$ cannot be performed analytically due to the (arbitrary) wave profile $A_{a}(x^-)$. 

Thus the final expression for the tree-level 3-gluon amplitude is 
\be\label{3p2}
A_{3}=2\im g\,f^{\sa_{1}\sa_{2}\sa_{3}}\,\delta^{d-1}\!\left(\sum_{r=1}^{3}k_{r}\right)\,\int \d x^{-}\left(\epsilon_{1}\cdot\epsilon_{2}\, K_{1}\cdot\epsilon_{3} + \mathrm{cyclic}\right)\,\exp\left[\im \sum_{s=1}^{3}\frac{f_{s}}{2\,k_{+\,s}}\right]\,,
\ee
where
\be\label{deltad}
\delta^{d-1}\!\left(\sum_{r=1}^{3}k_{r}\right):=\delta_{\left(\sum_{r=1}^{3}e_r\right),0}\; \delta\!\left(\sum_{r=1}^{3}k_{+\,r}\right)\,\delta^{d-2}\!\left(\sum_{r=1}^{3}\bk_{r}\right)\,,
\ee
and the functions $f_{s}(x^-)$ are defined by \eqref{phi} with $m=0$. In arriving at \eqref{3p2} from \eqref{3p1}, we have used the identities
\begin{eqnarray}
\label{gpolrels}
K_{{r}}\cdot \epsilon_{{s}}&=&\left\{\begin{array}{c c}
                                          0 & \mathrm{if}\;\; {r}={s} \\
             \frac{\epsilon_{s}^{a}}{k_{+\,s}}\,(k_{+\,r}\,k_{a\,s}-k_{+\,s}\,k_{a\,r}+ A_{a}(k_{+\,r}\, e_s- k_{+\,s}\,e_r)) \quad & \mathrm{otherwise}
                                         \end{array}\right. \,,\\
\label{gpolrels2}
\epsilon_{{r}}\cdot\epsilon_{{s}}&=&\left\{\begin{array}{c c}
                                          0 & \mathrm{if}\;\; {r}={s} \\
                                          -\epsilon_{a\,r}\,\epsilon_{s}^{a} & \mathrm{otherwise}
                                         \end{array}\right.\,.
\end{eqnarray}
In particular, the integrand has non-trivial functional dependence on $x^-$ through the `kinematics' as well as the overall exponential factor. In a flat background, it is easy to see that $x^-$-dependence drops out of the kinematics, and the $\d x^-$ integration against the exponential factor results in a momentum conserving delta function in the remaining light cone direction. Further, for the special case of the impulsive plane wave (where $A_{a}(x-)$ has delta function support), this 3-point amplitude has been evaluated explicitly~\cite{Adamo:2017nia}.


\subsection{4-points}

The tree-level 4-point amplitude receives contributions from three exchange diagrams as well as the 4-point contact interaction. In a flat background, the norm of the exchanged momenta are proportional to the Mandelstam invariants $\ms=k_{1}\cdot k_2$, $\st=k_{1}\cdot k_{3}$, $\su=k_{1}\cdot k_{4}$. On the plane wave background, these `invariants' are no longer straightforward numbers, since norms of exchanged momenta are now non-trivial functions of $x^-$. However, we can still use the Mandelstam labels to distinguish between the distinct exchange diagram contributions to the amplitude; this means that the 4-point gluon amplitude can be written as
\be\label{4p1}
A_{4}=A_{4}^{\mathrm{cont}}+A_{4,\ms}^{\mathrm{exch}}+A_{4,\st}^{\mathrm{exch}}+A_{4,\su}^{\mathrm{exch}}\,,
\ee
where $A_{4}^{\mathrm{cont}}$ is the contribution from the 4-point contact interaction and the $A_{4}^{\mathrm{exch}}$ are the exchange diagrams.

For simplicity, we normalise the amplitude by an overall factor of $\cN^{-1}_{(d)}$. The contact contribution is given by evaluating the 4-point vertex \eqref{4pv} on four on-shell states:
\begin{multline}\label{4pc1}
A_{4}^{\mathrm{cont}}=\frac{g^{2}}{\cN_{(d)}}\,\delta^{d-1}\!\left(\sum_{r=1}^{4}k_r\right) \left[f^{\sa_{1}\sa_{2}\sb}f^{\sa_{3}\sa_{4}\sb}\,(\epsilon_{1}\cdot\epsilon_{3}\,\epsilon_{2}\cdot\epsilon_{4}-\epsilon_{1}\cdot\epsilon_{4}\,\epsilon_{2}\cdot\epsilon_{3}) \right. \\
\left. + f^{\sa_{1}\sa_{3}\sb}f^{\sa_{2}\sa_{4}\sb}\,(\epsilon_{1}\cdot\epsilon_{2}\,\epsilon_{3}\cdot\epsilon_{4}-\epsilon_{1}\cdot\epsilon_{4}\,\epsilon_{2}\cdot\epsilon_{3})+f^{\sa_{1}\sa_{4}\sb}f^{\sa_{2}\sa_{3}\sb}\,(\epsilon_{1}\cdot\epsilon_{2}\,\epsilon_{3}\cdot\epsilon_{4}-\epsilon_{1}\cdot\epsilon_{3}\,\epsilon_{2}\cdot\epsilon_{4})\right] \\
\times \int \d x^{-}\,\exp\left[\im\,\sum_{r=1}^{4}\frac{f_{r}}{2\,k_{+\,r}}\right]\,,
\end{multline}
where the position space integrals over $\d x^{+}\d^{d-2}\mathbf{x}$ have been performed to give the delta functions. The exchange diagrams require an insertion of the Feynman propagator $\scG^{F}_{\mu\nu}$, resulting in an integration over two space-time points, $X$ and $Y$ (the functions $f_r$ are those defined in \eqref{phi}).

As is typical in the plane wave background, all integrals except those in the light cone directions $\d x^{-}$, $\d y^{-}$ can be performed explicitly. This means that we do not obtain the $k^-$ momentum conserving delta function, but the $k^-$ integral can nevertheless be performed as a contour integral as above to obtain \eqref{prop2} from \eqref{fprop1}.
In the $\ms$-channel, this gives
\begin{multline}\label{4pes}
 A_{4,\ms}^{\mathrm{exch}}=g^{2}\,\delta^{d-1}\!\left(\sum_{r=1}^{4}k_r\right)\,f^{\sa_{1}\sa_{2}\sb}f^{\sa_{3}\sa_{4}\sb}\,\int \d^{2}\mu[\ms]\,\left[\epsilon_{1}\cdot\epsilon_{2}\,(K_{1}-K_{2})^{\mu}\right. \\
 \left.+2\epsilon_{1}\cdot K_{2}\,\epsilon^{\mu}_{2}-2\epsilon_{2}\cdot K_{1}\,\epsilon_{1}^{\mu}\right](x^{-})\,\D^{\ms}_{\mu\nu}(x^{-},y^{-}) \\
 \times\left[\epsilon_{3}\cdot\epsilon_{4}\,(K_{4}-K_{3})^{\nu}-2\epsilon_{3}\cdot K_{4}\,\epsilon^{\nu}_{4}+2\epsilon_{4}\cdot K_{3}\,\epsilon_{3}^{\nu}\right](y^{-}) \; +(x^-\leftrightarrow y^-)\,,
\end{multline}
where the integral is taken with respect to a measure associated with the $\ms$-channel:
\begin{multline}\label{smeas}
 \d^{2}\mu[\ms]=\Theta(x^--y^-)\,\frac{\d x^{-}\,\d y^{-}}{(k_{1}+k_{2})_{+}}\,\exp\left[\im\left(\frac{f_{1}(x^-)}{2k_{+\,1}}+\frac{f_{2}(x^-)}{2k_{+\,2}}+\frac{f_{3}(y^-)}{2k_{+\,3}}+\frac{f_{4}(y^{-})}{2k_{+\,4}}\right.\right. \\
 \left.\left. -\frac{1}{2(k_{1}+k_{2})_{+}}\int_{y^-}^{x^-} \d s\,\left((\bk_{1}+\bk_{2})+(e_1+e_2)\mathbf{A}(s)\right)^{2}\right)\right]\,.
\end{multline}
This measure arises from the integrations $\d x^+\d^{d-2}\mathbf{x}$ and $\d y^+\d^{d-2}\mathbf{y}$ in the exchange diagram. The object $\D^{\ms}_{\mu\nu}(x^{-},y^{-})$ is defined by \eqref{fprop}, with the Mandelstam superscript identifying the exchanged momentum. In particular,
\be\label{sD}
\D^{\ms}_{\mu\nu}(x^{-},y^{-}):=\left(\begin{array}{ccc}
                                                                                                                                              0 & 0 & 1 \\
                                                                                                                                              0 & -\delta_{ab} & \frac{(e_{1}+e_2)\Delta A_{a}}{(k_1+k_2)_+} \\
                                                                                                                                              1 & -\frac{(e_1+e_2) \Delta A_{c}}{(k_1+k_2)_+} & \frac{(e_1+e_2)^2 \Delta \mathbf{A}^{2}}{2(k_1+k_2)^2_+}
                                                                                                                                             \end{array}\right)\,.
\ee
Finally, the integrand is symmetrized with respect to $x^-\leftrightarrow y^-$ due to the Feynman propagator, which includes both advanced and retarded contributions.

Contributions from the other two exchange diagrams have a similar structure:
\begin{multline}\label{4pet}
 A_{4,\st}^{\mathrm{exch}}=g^{2}\,\delta^{d-1}\!\left(\sum_{r=1}^{4}k_r\right)\,f^{\sa_{1}\sa_{3}\sb}f^{\sa_{2}\sa_{4}\sb}\,\int \d^{2}\mu[\st]\,\left[\epsilon_{1}\cdot\epsilon_{3}\,(K_{1}-K_{3})^{\mu}\right. \\
 \left.+2\epsilon_{1}\cdot K_{3}\,\epsilon^{\mu}_{3}-2\epsilon_{3}\cdot K_{1}\,\epsilon_{1}^{\mu}\right](x^{-})\,\D^{\st}_{\mu\nu}(x^{-},y^{-}) \\
 \times\left[\epsilon_{2}\cdot\epsilon_{4}\,(K_{4}-K_{2})^{\nu}-2\epsilon_{2}\cdot K_{4}\,\epsilon^{\nu}_{4}+2\epsilon_{4}\cdot K_{2}\,\epsilon_{2}^{\nu}\right](y^{-}) \; +(x^-\leftrightarrow y^-)\,,
\end{multline}
\begin{multline}\label{4peu}
 A_{4,\su}^{\mathrm{exch}}=g^{2}\,\delta^{d-1}\!\left(\sum_{r=1}^{4}k_r\right)\,f^{\sa_{1}\sa_{4}\sb}f^{\sa_{2}\sa_{3}\sb}\,\int \d^{2}\mu[\su]\,\left[\epsilon_{1}\cdot\epsilon_{4}\,(K_{1}-K_{4})^{\mu}\right. \\
 \left.+2\epsilon_{1}\cdot K_{4}\,\epsilon^{\mu}_{4}-2\epsilon_{4}\cdot K_{1}\,\epsilon_{1}^{\mu}\right](x^{-})\,\D^{\su}_{\mu\nu}(x^{-},y^{-}) \\
 \times\left[\epsilon_{2}\cdot\epsilon_{3}\,(K_{3}-K_{2})^{\nu}-2\epsilon_{2}\cdot K_{3}\,\epsilon^{\nu}_{3}+2\epsilon_{3}\cdot K_{2}\,\epsilon_{2}^{\nu}\right](y^{-}) \; +(x^-\leftrightarrow y^-)\,,
\end{multline}
with the respective integral measures given by
\begin{multline}\label{tmeas}
 \d^{2}\mu[\st]=\Theta(x^--y^-)\,\frac{\d x^{-}\,\d y^{-}}{(k_{1}+k_{3})_{+}}\,\exp\left[\im\left(\frac{f_{1}(x^-)}{2k_{+\,1}}+\frac{f_{2}(y^-)}{2k_{+\,2}}+\frac{f_{3}(x^-)}{2k_{+\,3}}+\frac{f_{4}(y^{-})}{2k_{+\,4}}\right.\right. \\
 \left.\left. -\frac{1}{2(k_{1}+k_{3})_{+}}\int_{y^-}^{x^-} \d s\,\left((\bk_{1}+\bk_{3})+(e_1+e_3)\mathbf{A}(s)\right)^{2}\right)\right]\,,
\end{multline}
and
\begin{multline}\label{umeas}
 \d^{2}\mu[\su]=\Theta(x^--y^-)\,\frac{\d x^{-}\,\d y^{-}}{(k_{1}+k_{4})_{+}}\,\exp\left[\im\left(\frac{f_{1}(x^-)}{2k_{+\,1}}+\frac{f_{2}(y^-)}{2k_{+\,2}}+\frac{f_{3}(y^-)}{2k_{+\,3}}+\frac{f_{4}(x^{-})}{2k_{+\,4}}\right.\right. \\
 \left.\left. -\frac{1}{2(k_{1}+k_{4})_{+}}\int_{y^-}^{x^-} \d s\,\left((\bk_{1}+\bk_{4})+(e_1+e_4)\mathbf{A}(s)\right)^{2}\right)\right]\,.
\end{multline}
This indicates that the entire 4-point tree amplitude is supported on $(d-1)$-dimensional momentum conservation as well as overall charge conservation (i.e., $\sum_{r=1}^{4}e_r=0$), as expected. It is straightforward to show that in a flat background the remaining light cone integrations can be performed analytically, resulting in the well-known 4-point gluon expression.


\section{Colour-Kinematics Duality}
\label{CKDual}

The Jacobi identity for structure constants of the gauge group ensures that there are linear relations between the colour structures contributing to every gluon amplitude beyond three points. In a flat perturbative background, \emph{colour-kinematics duality} is the statement that kinematic `numerators' of the amplitudes obey the same linear dependence relations as these colour structures~\cite{Bern:2008qj}. Around a flat background, it is well-known that the 4-point tree-level gluon amplitude can be written as
\be\label{fck1}
A_{4}|_{\mathrm{flat}}=g^2\,\delta^{d}\!\left(\sum_{r=1}^{4}k_r\right)\,\left(\frac{c_{\ms}\,N_{\ms}}{\ms}+\frac{c_{\st}\,N_{\st}}{\st}+\frac{c_{\su}\,N_{\su}}{\su}\right)\,,
\ee
where the $c$s are colour factors built from the structure constants, 
\be\label{colour}
c_{\ms}:=f^{\sa_{1}\sa_{2}\mathsf{b}}\,f^{\mathsf{b}\sa_{3}\sa_{4}}\,, \quad c_{\st}:=f^{\sa_{1}\sa_{3}\mathsf{b}}\,f^{\mathsf{b}\sa_{2}\sa_{4}}\,, \quad c_{\su}:=f^{\sa_{1}\sa_{4}\mathsf{b}}\,f^{\mathsf{b}\sa_{2}\sa_{3}}\,,
\ee
the $N$s are kinematic numerators and $\ms$, $\st$, $\su$ are the standard Mandelstam invariants. The only non-trivial step in obtaining this expression is the conversion of the contact contribution to the amplitude into a sum of exchange contributions, but this is easily accomplished by simply multiplying and dividing by the appropriate Mandelstam invariants.

The colour factors appearing in this amplitude are not linearly independent, due to the Jacobi identity
\be\label{cJac}
c_{\ms}-c_{\st}+c_{\su}=0\,.
\ee
Colour-kinematics duality for this amplitude is the statement:~\cite{Bern:2008qj}:
\be\label{kJac}
N_{\ms}-N_{\st}+N_{\su}=0\,,
\ee
namely, that the kinematic numerators obey the same Jacobi-like identity. Since the colour structure of the 4-point gluon amplitude is un-affected by the presence of a background plane wave gauge field, the colour Jacobi identity is un-changed. It is natural to ask: does the colour-kinematics duality continue to hold in the presence of the background field?

Obviously, \eqref{kJac} cannot be na\"ively generalized to the plane wave background: first of all, the 4-point amplitude cannot be written in the simple form \eqref{fck1} as there are integrals which cannot be performed analytically. In this sense, it is not even clear how to identify the analogues of `kinematic numerators' or indeed how to combine them in a sensible linear relation. Furthermore, there is only $(d-1)$-dimensional momentum conservation on the plane wave background. These factors make it difficult to see how any remnant of the duality can survive in the presence of background fields.

However, by identifying the appropriate curved background versions of the kinematic numerators and finding a suitable linear relation between them, a plane wave analogue of colour-kinematics duality can indeed be shown to exist. It necessarily reduces  to \eqref{kJac} in the flat background limit.


\subsection{Colour-kinematics representation}

A minimal criterion for obtaining a colour-kinematics representation (at 4-points) is the ability to express $A_4$ so that every contribution is associated with one of the Mandelstam labels. The contact contribution $A_{4}^{\mathrm{cont}}$ given by \eqref{4pc1} appears to present an obstruction, but this is overcome after some straightforward manipulations. In terms of the colour factors, $A_4^{\mathrm{cont}}$ can be written as:
\begin{multline}\label{cm1}
 \frac{g^{2}}{\cN_{(d)}}\, \left[c_{\ms}\,(\epsilon_{1}\cdot\epsilon_{3}\,\epsilon_{2}\cdot\epsilon_{4}-\epsilon_{1}\cdot\epsilon_{4}\,\epsilon_{2}\cdot\epsilon_{3}) + c_{\st}\,(\epsilon_{1}\cdot\epsilon_{2}\,\epsilon_{3}\cdot\epsilon_{4}-\epsilon_{1}\cdot\epsilon_{4}\,\epsilon_{2}\cdot\epsilon_{3})\right. \\
 \left.+c_{\su}\,(\epsilon_{1}\cdot\epsilon_{2}\,\epsilon_{3}\cdot\epsilon_{4}-\epsilon_{1}\cdot\epsilon_{3}\,\epsilon_{2}\cdot\epsilon_{4})\right] \int \d^{d}X\,\exp\left[\im\sum_{r=1}^{4}\phi_{r}(X)\right]\,,
\end{multline} 
which suggests that certain sets of terms should naturally be associated with each exchange channel.

To do this, the remaining integration must be cast in the form appearing for the exchange channels. Take the terms in \eqref{cm1} proportional to $c_{\ms}$:
\begin{multline}\label{cm2}
\frac{g^{2}c_{\ms}}{\cN_{(d)}}\,(\epsilon_{1}\cdot\epsilon_{3}\,\epsilon_{2}\cdot\epsilon_{4}-\epsilon_{1}\cdot\epsilon_{4}\,\epsilon_{2}\cdot\epsilon_{3}) \\
\times\,\int \d^{d} X\,\d^{d} Y\,\delta^{d}(X-Y)\,\exp\left[\im\sum_{r=1,2}\phi_{r}(X)+\im\sum_{s=3,4}\phi_{s}(Y)\right]\,,
\end{multline}
where an auxiliary integration over $\d^{d}Y$ has been inserted to make the position dependence of the exponential compatible with the $\ms$-channel. Now, the charged \emph{scalar} propagator,
\be\label{csprop}
\scG^{F}(X,Y)= \frac{\cN_{(d)}}{2\,\pi\,\im}\,\oint \frac{\d^{d}k}{k^{2}+\im\,\varepsilon}\, \exp\left[\im\phi_{k}(X)-\im\phi_{k}(Y)\right]\,,
\ee
obeys an equation of motion
\be\label{csprop2}
D^{\mu} D_{\mu}\,\scG^{F}(X,Y)=\delta^{d}(X-Y)=D^{\prime\mu} D^{\prime}_{\mu}\,\scG^{F}(X,Y)\,,
\ee
for $D_{\mu}$ the covariant derivative with respect to $X$ and $D^{\prime}_{\mu}$ the covariant derivative with respect to $Y$. This scalar propagator enables a trivial re-writing of \eqref{cm2}:
\begin{multline}\label{cm3}
 \frac{g^2 c_{\ms}}{2\,\cN_{(d)}}\,(\epsilon_{1}\cdot\epsilon_{3}\,\epsilon_{2}\cdot\epsilon_{4}-\epsilon_{1}\cdot\epsilon_{4}\,\epsilon_{2}\cdot\epsilon_{3}) \int \d^{d} X\,\d^{d} Y\,\left(D^{\mu} D_{\mu}\,\scG^{F}(X,Y) \right. \\
 \left.+D^{\prime\mu} D^{\prime}_{\mu}\,\scG^{F}(X,Y)\right)\,\exp\left[\im\sum_{r=1,2}\phi_{r}(X)+\im\sum_{s=3,4}\phi_{s}(Y)\right]\,,
\end{multline}
where the conversion into a wave operator acting on the propagator has been done symmetrically with respect to $X$ and $Y$ derivatives.

After integrating by parts twice on each term of the integrand and then performing as many integrations as possible, one finds:
\begin{multline}\label{cm4}
 -g^{2}\,\delta^{d-1}\!\left(\sum_{r=1}^{4}k_r\right)\, c_{\ms}\,(\epsilon_{1}\cdot\epsilon_{3}\,\epsilon_{2}\cdot\epsilon_{4}-\epsilon_{1}\cdot\epsilon_{4}\,\epsilon_{2}\cdot\epsilon_{3}) \\
 \times\,\int \d^{2}\mu[\ms]\,\left(K_{1}\cdot K_{2}(x^-)+K_{3}\cdot K_{4}(y^-)\right) \: +(x^{-}\leftrightarrow y^{-})\,, 
\end{multline}
which is now manifestly in a form compatible with the $\ms$-channel exchange contribution \eqref{4pes}. Similar manipulations can be performed on the terms in \eqref{cm1} proportional to $c_{\st}$ and $c_{\su}$, with the result that the 4-point amplitude is re-cast in the form:
\be\label{ckrep1}
A_{4}=g^{2}\,\delta^{d-1}\!\left(\sum_{r=1}^{4}k_r\right)\left(c_{\ms}\,\int\d^{2}\mu[\ms]\,n_{\ms}+c_{\st}\,\int\d^{2}\mu[\st]\,n_{\st}+c_{\su}\,\int\d^{2}\mu[\su]\,n_{\su}\right)\,,
\ee
where integrands associated to each channel are
\begin{multline}\label{ns}
 n_{\ms}=\left[\epsilon_{1}\cdot\epsilon_{2}\,(K_{1}-K_{2})^{\mu}+2\epsilon_{1}\cdot K_{2}\,\epsilon^{\mu}_{2}-2\epsilon_{2}\cdot K_{1}\,\epsilon_{1}^{\mu}\right](x^{-})\,\D^{\ms}_{\mu\nu}(x^{-},y^{-}) \\
 \times\left[\epsilon_{3}\cdot\epsilon_{4}\,(K_{4}-K_{3})^{\nu}-2\epsilon_{3}\cdot K_{4}\,\epsilon^{\nu}_{4}+2\epsilon_{4}\cdot K_{3}\,\epsilon_{3}^{\nu}\right](y^{-}) \\
 -(\epsilon_{1}\cdot\epsilon_{3}\,\epsilon_{2}\cdot\epsilon_{4}-\epsilon_{1}\cdot\epsilon_{4}\,\epsilon_{2}\cdot\epsilon_{3})\left(K_{1}\cdot K_{2}(x^-)+K_{3}\cdot K_{4}(y^-)\right)+(x^-\leftrightarrow y^-)\,,
\end{multline}
\begin{multline}\label{nt}
 n_{\st}=\left[\epsilon_{1}\cdot\epsilon_{3}\,(K_{1}-K_{3})^{\mu}+2\epsilon_{1}\cdot K_{3}\,\epsilon^{\mu}_{3}-2\epsilon_{3}\cdot K_{1}\,\epsilon_{1}^{\mu}\right](x^{-})\,\D^{\st}_{\mu\nu}(x^{-},y^{-}) \\
 \times\left[\epsilon_{2}\cdot\epsilon_{4}\,(K_{4}-K_{2})^{\nu}-2\epsilon_{2}\cdot K_{4}\,\epsilon^{\nu}_{4}+2\epsilon_{4}\cdot K_{2}\,\epsilon_{2}^{\nu}\right](y^{-}) \\
 -(\epsilon_{1}\cdot\epsilon_{2}\,\epsilon_{3}\cdot\epsilon_{4}-\epsilon_{1}\cdot\epsilon_{4}\,\epsilon_{2}\cdot\epsilon_{3})\left(K_{1}\cdot K_{3}(x^-)+K_{2}\cdot K_{4}(y^-)\right)+(x^-\leftrightarrow y^-)\,,
\end{multline}
\begin{multline}\label{nu}
 n_{\su}=\left[\epsilon_{1}\cdot\epsilon_{4}\,(K_{1}-K_{4})^{\mu}+2\epsilon_{1}\cdot K_{4}\,\epsilon^{\mu}_{4}-2\epsilon_{4}\cdot K_{1}\,\epsilon_{1}^{\mu}\right](x^{-})\,\D^{\su}_{\mu\nu}(x^{-},y^{-}) \\
 \times\left[\epsilon_{2}\cdot\epsilon_{3}\,(K_{3}-K_{2})^{\nu}-2\epsilon_{2}\cdot K_{3}\,\epsilon^{\nu}_{3}+2\epsilon_{3}\cdot K_{2}\,\epsilon_{2}^{\nu}\right](y^{-}) \\
 -(\epsilon_{1}\cdot\epsilon_{2}\,\epsilon_{3}\cdot\epsilon_{4}-\epsilon_{1}\cdot\epsilon_{3}\,\epsilon_{2}\cdot\epsilon_{4})\left(K_{1}\cdot K_{4}(x^-)+K_{2}\cdot K_{3}(y^-)\right)+(x^-\leftrightarrow y^-)\,.
\end{multline}
The representation \eqref{ckrep1} is the closest we can get to the colour-kinematics representation \eqref{fck1} for the 4-point gluon amplitude in a plane wave background. Indeed, in the flat background limit, it is easy to see that the two representations are equal; in particular
\be\label{flatlim}
\left.\int\d^{2}\mu[\ms]\,n_{\ms}\right|_{\mathbf{A}=\mathrm{const.}}=\delta\!\left(\sum_{r=1}^{4} k_{-\,r}\right)\,\frac{N_{\ms}}{\ms}\,,
\ee
and likewise for the other channels.


\subsection{Kinematic Jacobi identity}

In searching for a notion of colour-kinematics duality on a plane wave background, the key question is: what is a reasonable analogue of the kinematic Jacobi identity \eqref{kJac} in the background field? In a flat background, isolating the kinematic numerators entails stripping off overall momentum conserving delta functions and scalar propagators. In the plane wave background, there are $(d-1)$ such delta functions which can be trivially stripped off, but the ingredients which would have led to the $d^{\mathrm{th}}$-momentum conserving delta function as well as propagator factors in a flat background are now encoded in the complicated integral measures $\d^{2}\mu[\ms]$, etc.

This suggests that the appropriate analogues of the kinematic numerators $\{N_{\ms},N_{\st},N_{\su}\}$ in a plane wave are in fact the `tree-level integrands' $\{n_{\ms},n_{\st},n_{\su}\}$ defined by \eqref{ns} -- \eqref{nu}. Of course, these objects are highly gauge dependent (indeed, the gauge invariant amplitude is only obtained after summing their integrals), but then again so are the kinematic numerators in the flat background. Furthermore, \eqref{flatlim} demonstrates that isolating (for instance) $n_{\ms}$ is equivalent to isolating $N_{\ms}$ in the flat background limit. This notion of working with `tree-level integrands' was already seen to be useful when considering the double copy between 3-point gluon and graviton amplitudes on plane wave background~\cite{Adamo:2017nia}.

However, the $\{n_{\ms},n_{\st},n_{\su}\}$ are still not suitable to be combined in a linear relation akin to \eqref{kJac} due to the presence of $\D_{\mu\nu}(x^-,y^-)$ insertions. For example, isolating all terms proportional to $\epsilon_1\cdot\epsilon_2$ $\epsilon_{3}\cdot\epsilon_4$ in the tree-level integrands and combining them in a relation analogous to \eqref{cJac} leads to
\begin{multline}\label{wrong1}
\left(n_{\ms}-n_{\st}+n_{\su}\right)|_{\epsilon_1\cdot\epsilon_2\,\epsilon_{3}\cdot\epsilon_{4}}=(K_1-K_2)^{\mu}(x^-)\,\D^{\ms}_{\mu\nu}\,(K_{4}-K_3)^{\nu}(y^-) +K_{1}\cdot K_{3}(x^-) \\
+K_{2}\cdot K_{4}(y^-)-K_{1}\cdot K_{4}(x^-)+K_{2}\cdot K_{3}(y^-) \: +(x^- \leftrightarrow y^-)\,.
\end{multline}
Not only is this expression non-vanishing, it also seems to have no interesting structure -- other than vanishing in the flat background limit.

The problem lies with the insertions of $\D_{\mu\nu}$. These encode the tensor structure of the propagator (which we certainly want to include in any plane wave analogue of a kinematic numerator), but they also carry information akin to a scalar propagator. This is because $\D_{\mu\nu}$ `knows' about the exchanged momenta and charges associated to the channel in which it appears. For example, in the $\ms$-channel, $\D^{\ms}_{\mu\nu}$ contains terms proportional to
\be\label{spfac}
\frac{e_1+e_2}{(k_1+k_2)_{+}}\,.
\ee
Although such terms vanish in the flat background limit (they are always accompanied by factors of $\Delta\mathbf{A}$), their inclusion on a plane wave background is akin to including denominators associated with a scalar propagator. Indeed, it is precisely these terms which obstruct the combination $n_\ms-n_\st+n_\su$ from giving something nice, as demonstrated by \eqref{wrong1}. 

Resolving this problem boils down to setting all exchanged momenta/charge factors in $\D_{\mu\nu}$ equal across all channels \emph{before} forming any linear relation between the integrands. Intuitively, this can be thought of as `putting everything over a common denominator'; the flat background analogue would be to consider a kinematic Jacobi identity
\be\label{altkin}
\frac{N_{\ms}}{\ms}-\frac{N_{\st}}{\ms}+\frac{N_{\su}}{\ms}=0\,,
\ee
which is, of course, identical to \eqref{kJac}. To operationalize this procedure, select a reference channel: this can be any one of $\ms$, $\st$ or $\su$. Without loss of generality, we choose the $\ms$-channel. Define a map $\sigma$ which acts on the $\D_{\mu\nu}$ insertions as:
\be\label{submap}
\sigma(\D^{\ms}_{\mu\nu})=\D^{\ms}_{\mu\nu}\,, \qquad \sigma(\D^{\st}_{\mu\nu})=\D^{\ms}_{\mu\nu}\,, \qquad \sigma(\D^{\su}_{\mu\nu})=\D^{\ms}_{\mu\nu}\,,
\ee
setting all factors appearing in the integrands which are associated to exchanged momenta equal to \eqref{spfac}. In the flat background limit, $\sigma$ reduces to the identity map.

\medskip

This enables us to consider a linear combination of the tree-level integrands with relative signs chosen to mimic the colour Jacobi relation \eqref{cJac}, namely:
\be\label{kJac1}
\sigma\left(n_\ms-n_\st +n_\su\right)\,.
\ee
The various terms which enter this can be classified by their dependence on the particle polarizations; in particular, terms either go like $(\epsilon\cdot\epsilon)^2$ or $(\epsilon\cdot\epsilon)^1$. Overall covariance of the integrands ensures that considering particular representatives of each class captures the general behaviour. To this end, we can look explicitly at terms proportional to $\epsilon_{1}\cdot\epsilon_{2}\,\epsilon_{3}\cdot\epsilon_{4}$ as representatives of $(\epsilon\cdot\epsilon)^2$, and terms proportional to a single power of $\epsilon_{1}\cdot\epsilon_{2}$ as representatives of $(\epsilon\cdot\epsilon)^1$.

Adopting the notation
\be\label{propnote}
K_{i}^{\mu}(x^-)\,\D^{\ms}_{\mu\nu}(x^-,y^-)\,K_{j}^{\nu}(y^-):=(i|\ms|j)\,, \qquad \epsilon^{\mu}_{i}(x^-)\,\D^{\ms}_{\mu\nu}(x^-,y^-)\,K_{j}^{\nu}(y^-):=(\epsilon_i|\ms|j)\,,
\ee
for contractions between momenta or polarizations via the propagator tensor structure, all terms proportional to $\epsilon_{1}\cdot\epsilon_{2}\,\epsilon_{3}\cdot\epsilon_{4}$ in each of the integrands are given by:
\be\label{2es}
n_{\ms}|_{\epsilon_{1}\cdot\epsilon_{2}\,\epsilon_{3}\cdot\epsilon_{4}}=(1-2|\ms|4-3)\,,
\ee
\be\label{2et}
n_{\st}|_{\epsilon_{1}\cdot\epsilon_{2}\,\epsilon_{3}\cdot\epsilon_{4}}=-(1|\st|1+3)-(2|\st|2+4)\,,
\ee
\be\label{2eu}
n_{\su}|_{\epsilon_{1}\cdot\epsilon_{2}\,\epsilon_{3}\cdot\epsilon_{4}}=-(1|\su|1+4)-(2|\su|2+3)\,,
\ee
before symmetrizing over $x^-\leftrightarrow y^-$. In arriving at these expressions, one makes extensive use of the identity \eqref{pident}; for instance, the first $\st$-channel contribution \eqref{2et} is obtained from
\be\label{2et*}
K_{1}\cdot K_{3}(x^-)=K_{1}\cdot (K_{1}+K_{3})(x^-)=K_{1}^{\mu}(x^-)\,\D^{\st}_{\mu\nu}(x^-,y^-)\,(K_1+K_3)^{\nu}(y^-) = (1|\st|1+3)\,.
\ee
Symmetrizing over $x^-\leftrightarrow y^-$ and applying the map $\sigma$, one immediately discovers that
\be\label{2eck}
\sigma\left(n_{\ms}-n_{\st}+n_{\su}\right)|_{\epsilon_{1}\cdot\epsilon_{2}\,\epsilon_{3}\cdot\epsilon_{4}}=0\,.
\ee
So the $(\epsilon\cdot\epsilon)^2$ structures in the linear combination \eqref{kJac1} obey a Jacobi-like identity.

Observing that $(\epsilon_{i}|\ell|\epsilon_{j})=\epsilon_{i}\cdot\epsilon_{j}$ for any choice of channel $\ell=\ms,\st,\su$ and polarization vectors, it follows that all terms proportional to a single power of $\epsilon_{1}\cdot\epsilon_{2}$ are given in each tensor structure by:
\be\label{1es}
n_{\ms}|_{\epsilon_{1}\cdot\epsilon_{2}}=2\left[(3+4|\ms|\epsilon_{3})\,(2-1|\ms|\epsilon_{4})+(1-2|\ms|\epsilon_3)\,(3+4|\ms|\epsilon_4)\right]\,,
\ee
\be\label{1et}
n_{\st}|_{\epsilon_{1}\cdot\epsilon_{2}}=-4\,(\epsilon_{3}|\st|1+3)\,(2+4|\st|\epsilon_{4})\,,
\ee
\be\label{1eu}
n_{\su}|_{\epsilon_{1}\cdot\epsilon_{2}}=-4\,(2+3|\su|\epsilon_{3})\,(\epsilon_{4}|\su|1+4)\,,
\ee
before symmetrizing $x^-\leftrightarrow y^-$. Note that the position dependence of the polarization vectors differs between the channels: in the $\ms$-channel, all polarization vectors are evaluated at $y^-$ whereas in the $\st$ and $\su$ channels the dependence is split. It is useful to arrange the contributions from all channels so that (before symmetrization) the position dependence of the polarizations is uniform; this leads to a re-expression of the contributions from the $\st$- and $\su$-channels:
\be\label{1et*}
n_{\st}|_{\epsilon_{1}\cdot\epsilon_{2}}=-4\,(1+3|\st|\epsilon_{3})\,(2+4|\st|\epsilon_{4})-4\,(2+4|\st|\epsilon_{4})\,\frac{\epsilon_{3}\cdot\Delta\mathbf{A}}{k_{+\,3}}(k_{+\,1}e_{3}-k_{+\,3}e_{1})\,,
\ee
\be\label{1eu*}
n_{\su}|_{\epsilon_{1}\cdot\epsilon_{2}}=-4\,(2+3|\su|\epsilon_{3})\,(1+4|\su|\epsilon_{4})-4\,(2+3|\su|\epsilon_{3})\,\frac{\epsilon_{4}\cdot\Delta\mathbf{A}}{k_{+\,4}}(k_{+\,1}e_{4}-k_{+\,4}e_{1})\,.
\ee
Upon symmetrizing with respect to $x^-\leftrightarrow y^-$, it follows that
\begin{multline}\label{1ess}
 n_{\ms}|_{\epsilon_{1}\cdot\epsilon_{2}}=(3+4|\ms|\epsilon_{3})\,(2-1|\ms|\epsilon_{4})+(1-2|\ms|\epsilon_3)\,(3+4|\ms|\epsilon_4) \\
+(\epsilon_{3}|\ms|3+4)\,(\epsilon_{4}|\ms|2-1) + (\epsilon_3|\ms|1-2)\,(\epsilon_4|\ms|3+4)
\end{multline}
\be\label{1ets}
n_{\st}|_{\epsilon_{1}\cdot\epsilon_{2}}=-2\left[(1+3|\st|\epsilon_{3})\,(2+4|\st|\epsilon_{4})+(\epsilon_{3}|\st|1+3)\,(\epsilon_4|\st|2+4)\right] +2\frac{\epsilon_{3}\cdot\Delta\mathbf{A}\,\epsilon_{4}\cdot\Delta\mathbf{A}}{k_{+\,3}\,k_{+\,4}}\,\mathrm{p}^{\st}\,,
\ee
\be\label{1eus}
n_{\su}|_{\epsilon_{1}\cdot\epsilon_{2}}=-2\left[(2+3|\su|\epsilon_{3})\,(1+4|\su|\epsilon_{4})+(\epsilon_{3}|\su|2+3)\,(\epsilon_4|\su|1+4)\right] +2\frac{\epsilon_{3}\cdot\Delta\mathbf{A}\,\epsilon_{4}\cdot\Delta\mathbf{A}}{k_{+\,3}\,k_{+\,4}}\,\mathrm{p}^{\su}\,,
\ee
where $\mathrm{p}^{\ell}$ is a polynomial defined for each channel $\ell=\ms,\st,\su$ as
\be\label{poly}
\mathrm{p}^{\ms}=(k_{+\,1}e_2-k_{+\,2}e_1)(k_{+\,3}e_4-k_{+\,4}e_3)\,, \qquad \mathrm{p}^{\st}=(k_{+\,1}e_3-k_{+\,3}e_1)(k_{+\,2}e_4-k_{+\,4}e_2)\,, 
\ee
\begin{equation*}
\mathrm{p}^{\su}=(k_{+\,1}e_4-k_{+\,4}e_1)(k_{+\,2}e_3-k_{+\,3}e_2)\,.
\end{equation*}

It is easy to see that the map $\sigma$ acts on these polynomials in the fashion
\be\label{submap2}
\sigma(\mathrm{p}^{\ms})=\mathrm{p}^{\ms}\,, \qquad \sigma(\mathrm{p}^{\st})=\mathrm{p}^{\ms}\,, \qquad \sigma(\mathrm{p}^{\su})=\mathrm{p}^{\ms}\,,
\ee
since the polynomials themselves arise from manipulations of terms which are constructed from $\D_{\mu\nu}$ insertions appropriate to each channel. This leaves us with all ingredients necessary to consider the contributions of terms proportional to a single power of $\epsilon_{1}\cdot\epsilon_{2}$ in the linear relation \eqref{kJac1}. In doing so, a rather interesting fact is essential: despite the plane wave background resulting in reduced momentum conservation, momenta contracted with a polarization (either through the metric or $\D_{\mu\nu}$) act \emph{as if they were conserved}. This is because the polarizations project out any components of momentum in the $x^-$ lightcone direction -- which is precisely the direction in which momentum is not conserved. As a result, identities such as
\be\label{emomcon}
(1+2|\ms|\epsilon_{i})=-(3+4|\ms|\epsilon_i)\,,
\ee
are true for any $i=1,2,3,4$.

After some algebraic manipulations (on the support of $(d-1)$-dimensional momentum conservation), the final result is:
\begin{multline}\label{1eck}
 \sigma\left(n_{\ms}-n_{\st}+n_{\su}\right)|_{\epsilon_{1}\cdot\epsilon_{2}}=(3|\ms|\epsilon_3)\,(2-1|\ms|\epsilon_4)+(\epsilon_{3}|\ms|3)\,(\epsilon_4|\ms|2-1) \\
 +(1-2|\ms|\epsilon_3)\,(4|\ms|\epsilon_4)+(\epsilon_3|\ms|1-2)\,(\epsilon_4|\ms|4)\,.
\end{multline}
Although this is is non-zero, it does have a highly constraining structure. In particular, every term in \eqref{1eck} is proportional to $(i|\ms|\epsilon_{i})$, which reduces to $K_{i}\cdot\epsilon_{i}=0$ in a flat background. Thus, the Jacobi identity is obstructed by what could be referred to as `deformed gauge conditions': $(i|\ms|\epsilon_i)$ is deformed away from being pure gauge by the dependence of $\D_{\mu\nu}$ on the background field.

It is easy to see that the same thing happens for all contributions to \eqref{kJac1} proportional to a single power of $\epsilon\cdot\epsilon$. Define a congruence relation $\simeq$ by `equal up to deformed gauge conditions'; then the full statement of colour-kinematics duality on a plane wave background is that the kinematic integrands obey a Jacobi relation:
\be\label{kJac2}
\sigma\left(n_{\ms}-n_{\st}+n_{\su}\right)\simeq 0\,.
\ee
It should be emphasized that this relationship reduces to the standard kinematic Jacobi identity \eqref{kJac} in the flat background limit.


\section{Further Directions}
\label{FD}

By studying tree-level, four-gluon scattering on a sandwich plane wave background, we have demonstrated that a generalization of colour-kinematics duality persists in the presence of background curvature. There may well be some alternative expressions of colour-kinematics duality on these backgrounds,  but the key test will be whether they can play a significant  role in a curved background double copy construction. There are many other extensions and applications of this result which should also be explored in the future, but we briefly discuss four of them here: double copy, higher points, loops, and ambitwistor strings.


\subsection{Double copy}

The great utility of colour-kinematics duality on a flat perturbative background is the ability to trivially generate gravition scattering amplitudes at a given particle and loop order once the gauge theory amplitude of the same order is known in a colour-kinematics representation~\cite{Bern:2008qj,Bern:2010ue,Bern:2010yg}. Indeed, in such a representation one only needs to replace the colour structures of the gauge theory amplitude with another copy of the kinematic numerators to obtain the gravitational scattering amplitude. This double copy procedure reduces the computation of scattering amplitudes in perturbative gravity to finding a colour-kinematics representation for their gauge theory counterparts (at the level of the integrand). 

It is a theorem that colour-kinematics representations can always be found at tree-level~\cite{BjerrumBohr:2009rd,Stieberger:2009hq,BjerrumBohr:2010zs,Feng:2010my,Tye:2010dd}, and there is remarkable evidence in favour of their existence at higher loops as well~\cite{Carrasco:2011mn,Bern:2013yya,Bjerrum-Bohr:2013iza,Tourkine:2016bak,He:2016mzd,Hohenegger:2017kqy,Bern:2017yxu,He:2017spx,Geyer:2017ela}. Combined with double copy, this has enabled remarkable calculations in perturbative supergravity, up to \emph{five loops} with four external gluons at the time of writing~\cite{Bern:2017ucb,Bern:2018jmv}. However, it is not at all clear how -- or if -- double copy should generalize to gauge theory and gravity on non-trivial perturbative backgrounds despite the complete tree-level results.

In~\cite{Adamo:2017nia} it was shown that a double copy relationship between gauge theory and gravity exists at the level of three-point amplitudes on plane wave backgrounds. This relationship is more intricate than the simple squaring relation on a flat background, and includes certain replacement rules for converting the background gauge field of the gluon amplitude into the background metric of the graviton amplitude. The results of this paper suggest a natural way to proceed with double copy at four-points: namely, by replacing the colour factors $\{c_{\ms}, c_{\st}, c_{\su}\}$ appearing in \eqref{ckrep1} with another copy of the (suitably integrated) tree-level integrands $\{n_{\ms}, n_{\st}, n_{\su}\}$. After implementing the replacement rules for the background fields, we conjecture that such a procedure will generate the four-point graviton amplitude on a plane wave space-time. Of course, to check this one must first compute this graviton amplitude explicitly; this is the subject of on-going work.


\subsection{Higher points}

Using integration by parts arguments it is clear that the $n$-point tree-level gluon amplitudes on the plane wave background can always be put into the form
\be\label{npt}
A_{n}=g^{n-2}\,\delta^{d-1}\!\left(\sum_{r=1}^{n}k_r\right) \,\sum_{\Gamma\in\mathrm{cubic}}c_{\Gamma}\,\int \d\mu_{\Gamma}\, n_{\Gamma}\,,
\ee
where the sum is over the $(2n-5)!!$ possible cubic diagrams, and $c_{\Gamma}$, $\d\mu_{\Gamma}$, $n_\Gamma$ are the associated colour factors, $(n-2)$-dimensional measures, and tree-level integrands, respectively. Due to the Jacobi identity, only $(n-2)!$ of the colour factors are independent, so the statement of colour-kinematics duality becomes
\be\label{nJac}
c_{\alpha}-c_{\beta}+c_{\gamma}=0 \iff \sigma\left(n_{\alpha}-n_{\beta}+n_{\gamma}\right)\simeq 0\,,
\ee
for any cubic graphs $\alpha,\beta,\gamma$ whose colour factors are related by this Jacobi identity.

In the four-point case studied here, the assignment of contact terms (i.e., the contribution of the 4-point vertex) to each of the three exchange channels was straightforward as there are only three colour structures and one Jacobi identity between them. However, at higher points this is not the case: the five-gluon tree amplitude has 15 independent exchange structures (each with two propagators) but only six independent colour structures~\cite{Bern:2008qj}. This makes the assignment of contributions from contact terms to each exchange structure much more non-trivial.

On a flat background, the crucial guiding structures are the kinematic Jacobi identities themselves. In particular, the requirement of colour-kinematics duality non-trivially constrains the representation of the amplitudes at higher points: \emph{not} every way of breaking apart the contact terms consistent with the colour structures does obey colour-kinematics duality. Explicit solutions for the appropriate kinematic numerators have now been given using a variety of different methods (e.g., \cite{BjerrumBohr:2010zs,Mafra:2011kj}).

On a plane wave background, the colour structures and Jacobi identities relating them will be the same as in flat space. By virtue of the fact that \eqref{nJac} reduces to the identity $N_{\alpha}-N_{\beta}+N_{\gamma}=0$ for kinematic numerators in the flat background limit, it is clear that the plane wave colour-kinematics duality non-trivially constrains the representations of amplitudes beyond four-points.

It would be interesting to see how this works explicitly, even at five-points. All contributions to the tree-level five-gluon amplitude are proportional to one or two powers of $(\epsilon\cdot\epsilon)$. At four points, those terms proportional to $(\epsilon\cdot\epsilon)^2$ obeyed the kinematic Jacobi identity with an exact equality, while those proportional to only a single power of $(\epsilon\cdot\epsilon)$ obeyed the identity only up to the $\simeq$ relation. Does the same thing happen at five points? 


\subsection{Loops}

\begin{figure}[t]
\centering
\includegraphics[scale=.6]{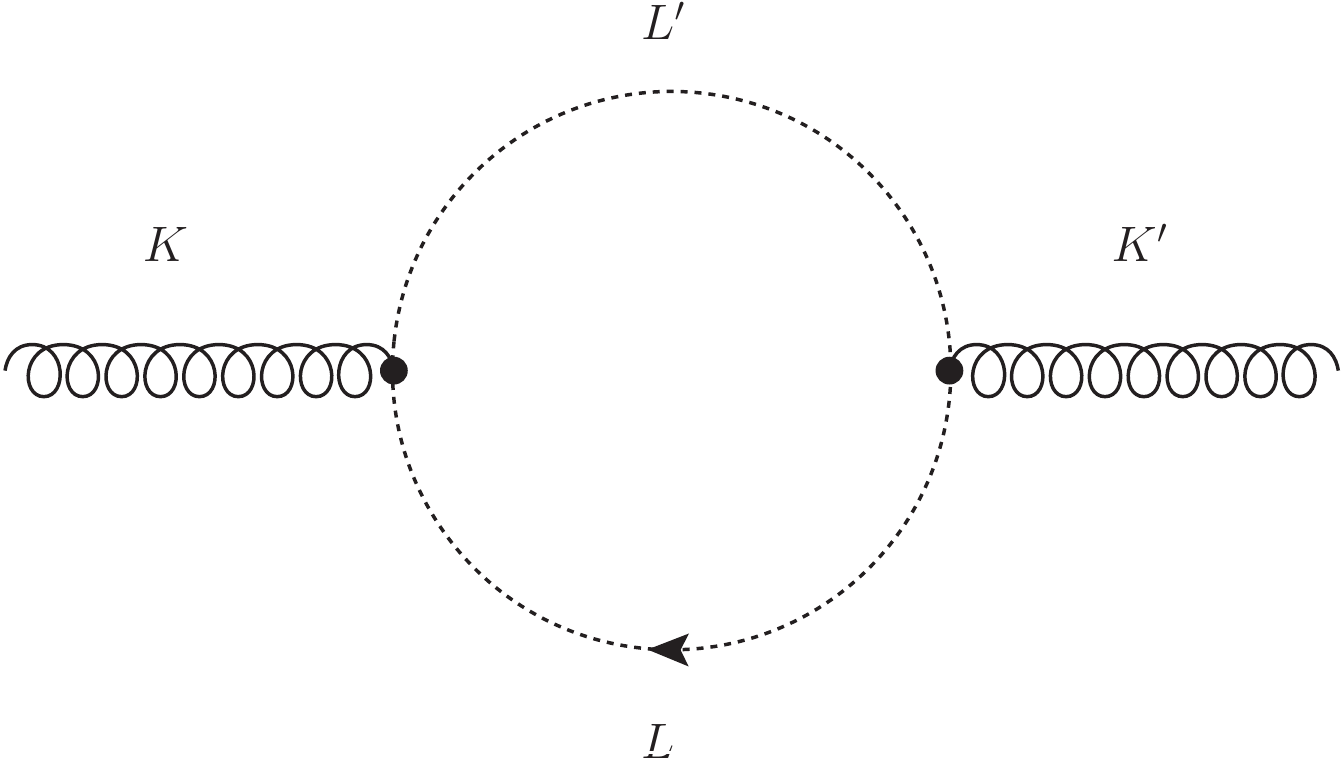}
\caption{The ghost loop contribution to the gluon self-energy}
\label{Ghostloop}
\end{figure}

Having determined the complete set of Feynman rules for gauge theory on a plane wave background, computations at \emph{all} orders in perturbation theory are, at least in principle, possible. It would be intriguing to apply these Feynman rules at loop level. In the spirit of this paper, one interesting question would be the fate of colour-kinematics duality for the one-loop, four-point gluon amplitude. But on an even  more basic level, one could consider one-loop corrections to the gluon self-energy. As on a flat background, this receives contributions from three different Feynman diagrams: the gluon loop, gluon tadpole, and ghost loop. Of these three, the ghost loop (see Figure \ref{Ghostloop}) is the easiest since the scalar ghosts in the loop produce the simplest tensor structure. 

Let us see how far we can get trying to evaluate this particular Feynman diagram in a plane wave background. The usual subtleties associated with the background mean that \emph{a priori} we can make no assumptions about the relationships between the momenta $K,K^{\prime},L, L^{\prime}$ without due care. The most basic expression for the ghost loop is therefore:
\begin{multline}\label{gl1}
g^{2}\,C_{2}(G)\,\delta^{\sa\sb}\,\int \d^{d}X\,\d^{d}Y\,\frac{\d^{d}l}{l^2+\im\,\varepsilon}\,\frac{\d^{d}l^{\prime}}{(l^{\prime})^2+\im\,\varepsilon}\,L^{\prime\,\mu}(x^{-})\,L^{\nu}(y^-) \\
\times \exp\left[\im\left(\phi_{k}(X)+\phi_{k'}(Y)+\phi_{l}(X)-\phi_{l}(Y)+\phi_{l'}(Y)-\phi_{l'}(X)\right)\right]\,,
\end{multline}
where $C_{2}(G)$ is the quadratic Casimir of the gauge group. Performing all the straightforward integrations in this expression gives various relations:
\be\label{gl2}
 e'=-e\,, \quad e_{l'}=e+e_l\,, \quad k^{\prime}_{+}=-k_{+}\,, \quad \bk^\prime=-\bk\,,
\ee
\begin{equation*}
l^{\prime}_{+}=(k+l)_{+}\,, \quad \mathbf{l}^{\prime}=\bk+\mathbf{l}\,, \quad l_{-}=\frac{\mathbf{l}^2}{2\,l_+}\,, \quad l_{-}^{\prime}=\frac{(\bk+\mathbf{l})^2}{2\,(k+l)_+}\,,
\end{equation*}  
and leaves
\begin{multline}\label{gl3}
g^{2}\,C_{2}(G)\,\delta^{\sa\sb}\,\int \frac{\d x^{-}\,\d y^{-}\,\d l_{+}\,\d^{d-2}\mathbf{l}}{l_{+}\,(k+l)_{+}}\,L^{\prime\,\mu}(x^{-})\,L^{\nu}(y^-)\,\Theta(x^{-}-y^{-})\\
\times \exp\left[\im\left(k_{-}x^{-}+k_{-}^{\prime} y^{-}+\frac{1}{2\,k_+}\int^{x^-}_{y^-}\!\!\d s\,\left(2e\,\bk\cdot\mathbf{A}(s)+e^2\,\mathbf{A}^2(s)\right) \right.\right. \\
\left.\left. +\frac{1}{2\,l_+}\int^{x^-}_{y^-}\!\!\!\d s\,\left(\mathbf{l}+e_{l}\,\mathbf{A}(s)\right)^{2}-\frac{1}{2\,(k+l)_+}\int^{x^-}_{y^-}\!\!\!\d s\,\left(\bk+\mathbf{l}+(e+e_{l}) \mathbf{A}(s)\right)^{2}\right)\right] \,,
\end{multline}
plus another contribution with support on $y^- >x^-$.

It is easy to see that in the flat background limit, the remaining two light cone position integrals can be performed analytically, and the result is equal to the usual momentum space Feynman rules expression for the ghost loop in flat space. As is well-known, the flat-background contains UV divergences in the loop momentum integral and it seems that the same divergences underlie the plane wave expression \eqref{gl3}.

In the simpler (but related) context of QED on a plane wave background, the one-loop correction to the photon self-energy has been computed (for both on- and off-shell photons) in~\cite{Narozhny:1969,Dinu:2013gaa,Meuren:2014uia}, and divergences can be suitably regularized using `transverse dimensional regularization'~\cite{Casher:1976ae,Ilderton:2012qe}. This scheme regularizes divergences through the number of transverse dimensions in the light cone coordinates (i.e., $\d^{d-2}\mathbf{l}\rightarrow \d^{d-2-\epsilon}\mathbf{l}$). It would be interesting to see how far these techniques from QED can be pushed for Yang-Mills theory, with a view to fully regularizing \eqref{gl3} as well as the other diagrams contributing to the gluon self-energy at one loop     


\subsection{Ambitwistor strings}

An additional motivation for computing the 4-point gluon amplitude on a plane wave background is to provide `theoretical data' against which other methods for performing perturbative calculations in gauge theory can be checked. One example of such methods are \emph{ambitwistor strings}~\cite{Mason:2013sva}, a class of worldsheet models which describe field theories. On flat perturbative backgrounds, ambitwistor strings underpin the remarkable scattering equation formulae for the tree-level S-matrix of a wide array of massless QFTs~\cite{Cachazo:2013hca,Cachazo:2014xea}. Furthermore, coupling ambitwistor strings to curved (gauge or gravitational) background fields imposes the non-linear field equations on the background fields exactly, without any recourse to background perturbation theory~\cite{Adamo:2014wea,Adamo:2018hzd}.

Combined with the form of the worldsheet vertex operators on any such background~\cite{Adamo:2018ege}, this raises the possibility that ambitwistor strings could shed new light on observables in QFT on curved perturbative backgrounds. Indeed, we already showed that the 3-point tree-level amplitudes \eqref{3p2} can be obtained from a genus zero, three point correlation function in the heterotic ambitwistor string coupled to a background plane wave gauge field~\cite{Adamo:2017sze}. However, going to $n>3$ points in the ambitwistor string requires new insights: the analogue of the scattering equations on a curved background is not known, or equivalently, there are moduli integrals on curved backgrounds  which cannot be performed trivially.

Nevertheless, there are hints that the worldsheet may reproduce the 4-point amplitude \eqref{ckrep1} if these moduli integrals are handled appropriately, see~\cite{Ohmori:2015sha,Casali:2016atr,Casali:2017zkz}. On a flat background, ambitwistor strings result in $n-3$ moduli integrals for $n$ vertex insertions at genus zero; these integrals are performed against delta functions which impose the scattering equations. On a curved background, these $n-3$ moduli integrals can no longer be straightforwardly evaluated against delta functions, and in the specific case of a plane wave background there will be an additional zero mode integral (corresponding to the light cone coordinate $x^-$) which cannot be analytically performed. But this leaves $n-2$ integrals, which matches exactly the counting for a $n$-point tree-level scattering amplitude (expanded in cubic graphs) in a plane wave background.  

\acknowledgments

We thank Thomas Heinzl and Anton Ilderton for helpful comments on a draft. TA is supported by an Imperial College Junior Research Fellowship; LJM and EC were supported by EPSRC grant EP/ M018911/1; SN is supported by EPSRC grant EP/M50659X/1 and a Studienstiftung des deutschen Volkes scholarship. This research is supported in part by U.S. Department of Energy grant DE-SC0009999 and by funds provided by the University of California. This research was supported also in part by the National Science Foundation under Grant No.\ NSF PHY17-48958.

\bibliography{biblio}  
\bibliographystyle{JHEP}

\end{document}